\documentclass[twocolumn]{aastex7}

\usepackage{amsmath}
\usepackage{bm}
\usepackage[T1]{fontenc}

\graphicspath{{./}{figures/}}
\newcommand{\ts}{\textsuperscript}
\newcommand{\XCO}{X_\text{CO}}
\newcommand{\XCOi}{X_\text{CO,0}}

\begin{document}

\title{
Dust Recycling and Icy Volatile Enhancement (DRIVE): A Novel Method of Volatile Enrichment in Cold Giant Planets}

\author[0000-0002-5954-6302]{Eric R. Van Clepper}
\email{ericvc@uchicago.edu}
\affiliation{Department of the Geophysical Sciences, University of Chicago, 5734 S. Ellis Avenue, Chicago, IL 60637, USA}

\author[0000-0002-2692-7862]{Felipe Alarcón}
\email{felipe.alarcon@unimi.it}
\affiliation{Dipartimento di Fisica, Università degli Studi di Milano, Via Celoria 16, 20133 Milano, Italy}

\author[0000-0003-4179-6394]{Edwin Bergin}
\email{ebergin@umich.edu}
\affiliation{Department of Astronomy, University of Michigan, 323 West Hall, 1085 S. University Avenue, Ann Arbor, MI 48109, USA}

\author[0000-0002-0093-065X]{Fred J. Ciesla}
\email{fciesla@uchicago.edu}
\affiliation{Department of the Geophysical Sciences, University of Chicago, 5734 S. Ellis Avenue, Chicago, IL 60637, USA}

\received{2025 September 9}
\revised{2025 November 7}
\accepted{2025 November 10}
\submitjournal{ApJL}

\shorttitle{Dust Recycling and Icy Volatile Enhancement}
\shortauthors{Van Clepper et al.}

\begin{abstract}
\noindent

Giant planet atmospheres are thought to reflect the gas phase composition of the disk when and where they formed. However, these atmospheres may also be polluted via solid accretion or ice sublimation in the disk. Here, we propose a novel mechanism for enriching the atmospheres of these giant planets with volatiles via pebble drift, fragmentation, and ice sublimation. We use a combination of 3D hydrodynamic simulations, radiative transfer, and particle tracking to follow the trajectories and resulting temperatures of solids in a disk containing an embedded planet forming outside the CO snowline. We show that small dust can become entrained in the meridional flows created by the giant planet and advected above the disk midplane where temperatures are well above the sublimation temperature of CO. This transport of small grains occurs over 10 kyr timescales, with individual micron-sized grains cycling between the midplane and surface of the disk multiple times throughout the planetary accretion stage. We find that this stirring of dust results in sublimation of CO gas above the snow surface in the dust trap created exterior to the giant planet, leading to super-solar CO abundances in the pressure bump. This mechanism of Dust Recycling and Icy Volatile Enhancement in cold giant planets, which we call the DRIVE effect, may explain enhanced metallicities of both wide separation exoplanets as well as Jupiter in our own Solar System.

\end{abstract}

%%% Check the keywords!

\section{Introduction} \label{sec:intro}

A major goal of studying protoplanetary disk (PPD) evolution is to connect observed exoplanets with their formation regions within the disk.
Observations of young giant planets and PPDs, however, have revealed an unexplained mismatch between disk and planetary compositions.
Cold giant planets have been observed to have an enriched metallicity compared to their host star, both in our solar system and around other stars \citep[][]{atreya_deep_2020, li_water_2020, hsu_pds_2024, nasedkin_four---kind_2024, bergin_co_2024, balmer_jwst-tst_2025,  lothringer_refractory_2025}. The gas phase metallicity in the outer disk, however, should be low, with the majority of major C- and O-bearing molecules frozen out onto grains \citep{oberg_effects_2011, bergin_co_2024}. 
While the gas in the cold outer regions of PPDs is carbon rich and low metallicity, the atmospheres of giant planets that are inferred to form in these regions of the disk have lower C/O and higher metallicity than their host star.
Understanding the composition of the disk gas during the formation of these giant planets, and possible enrichment in volatile elements, is necessary for connecting observed compositions and metallicities of giant planets to the disk conditions in which they form.

One of the current leading theories for giant planet formation is pebble accretion, wherein planetary cores grow by accreting drifting pebbles through the disk \citep{lambrechts_forming_2014}. 
In this model, planetary core composition is set by the composition of these solids, both the refractory grain as well as any volatiles that may be present as an ice mantle.
Once the core grows large enough, it opens a gap in the disk, halting the inward drift of solids and shutting off pebble accretion \citep{lambrechts_separating_2014}. Though the pebble isolation mass depends on the gas scale height and turbulence, typical masses are on the order of a few 10s of Earth Masses \citep{bitsch_pebble-isolation_2018}. At this stage, the core accretes gas from the disk, undergoing a runaway growth phase until the gas disk dissipates \citep{pollack_formation_1996}. 
During this runaway growth phase, the atmosphere of the planet should accrete primarily gas, with the planet becoming more metal poor as it grows in mass, reflecting the metallicity of the PPD gas \citep{thorngren_MASS_2016}.
So while the core of the planet may initially be metal rich, once the planet reaches the pebble isolation mass the atmosphere should largely reflect the relatively metal poor composition of the gas where the planet formed.

One way to alter the composition of a giant planet atmosphere is if the planet grows inside of a snowline in the disk.
As pebbles drift from the outer disk, they will pass snowlines, enriching the region interior in volatiles \citep{cuzzi_material_2004, ciesla_evolution_2006,oberg_excess_2016, booth_chemical_2017, danti_composition_2023}. If a planet grows just interior to one of these snowlines, then its atmosphere should become enriched in those volatiles \citep{schneider_how_2021-1, schneider_how_2021}. Additionally, if the ice present on these grains carries with it trapped noble gasses, this sublimation may explain the enrichment of both volatiles and noble gasses in Jupiter's atmosphere \citep{monga_external_2014, mousis_jupiters_2019, atreya_deep_2020}. This model is particularly attractive as snowlines can also create local pebble pile-ups, which may help to initiate core formation \citep{drazkowska_planetesimal_2017, andama_role_2022, lau_rapid_2022}. This, however, requires the planetary core to accrete gas from the region just interior to the snowline, and cannot explain any enrichment of a given species exterior to its respective snowline.

Recent models have shown that the growing planetary embryo, surrounding gas, and smaller solids all interact with one another, complicating the otherwise simple picture of giant planet formation. One such complication is the meridional flow patterns of gas accretion onto the planet in 3D \citep{kley_three-dimensional_2001, morbidelli_meridional_2014, szulagyi_accretion_2014, fung_gap_2016, teague_meridional_2019,lega_gas_2024}. Hydrodynamic simulations of gap opening planets in 3D show that gas moves outward, away from the planet along the spiral arms at the midplane. Gas flows are away from the midplane at the gap edges and fall onto the poles of the planet from 1-3 scale heights above the midplane. Such 3D stirring can entrain small dust, increasing the dust scale height near the planet \citep{bi_puffed-up_2021} and even lead to the accretion of small dust onto the planet \citep{szulagyi_meridional_2022, maeda_delivery_2024, petrovic_material_2024, van_clepper_three-dimensional_2025}.

In this work, we examine the chemical consequences of the vertical stirring immediately outside a growing planet, and the delivery of volatiles to the planetary atmosphere.
First, we describe our methodological approach, including disk hydrodynamic simulations, radiative transfer modeling, Monte-Carlo particle integrations, and dust growth and fragmentation simulations. Using the results of these combined modeling techniques, we show that the Dust Recycling and Icy Volatile Enhancement (DRIVE) effect can lead to an enhancement of the gas phase abundance of volatiles, even exterior to the their respective snowlines.
Here, we use the enrichement of CO exterior to the CO snowline as an example, though the DRIVE effect can work in general, and is expected to function in a similar manner for other volatile ice species.
We discuss these results in the context of Jupiter and other directly imaged exoplanets and compare with other mechanisms of volatile enhancement in giant planet atmospheres.

\begin{figure*}
    \centering
    \includegraphics[width=\linewidth]{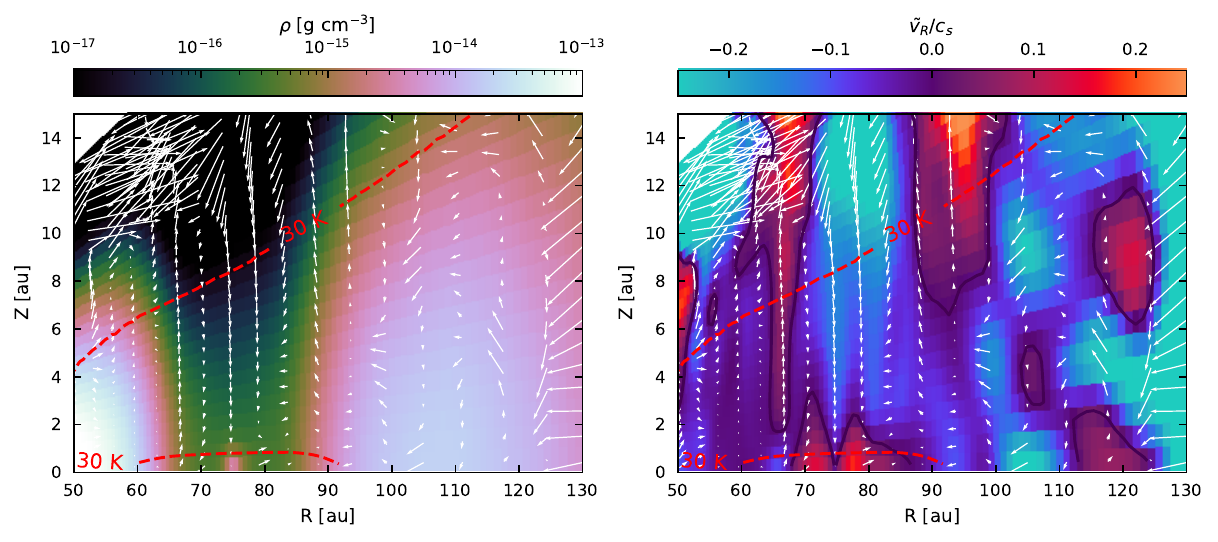}
    \caption{Azimuthally averaged gas density (left) and gas radial velocity relative to the planet (right) for our FARGO3D disk model containing a Jupiter mass planet at 75 au. The CO sublimation surface at 30 K recovered from our RADMC-3D model is indicated in both panels by the red dashed line. The region near the embedded planet is heated from a combination of accretion heating onto the planet, disk viscous heating, and radiative heating from the star, resulting in higher temperatures at the midplane in the gap. Gas advection relative to the local sound speed is also indicated by white arrows. Gas advection is primarily outward along the midplane near the planet, with significant vertical flows away from the midplane at the gap edge and in the pressure bump at 100 au.}
    \label{fig:disk_structure}
\end{figure*}

\section{Methodology}
\label{sec:methods}

To explore the consequences of dust stirring near an embedded giant planet, we model the gas disk using single-fluid FARGO3D hydrodynamic simulations \citep{masset_fargo_2000, benitez-llambay_fargo3d_2016} with 256 radial cells ranging from 30 au to 337.5 au, 512 cells in azimuth from 0 to 2$\pi$, and 32 cells in colatitude spanning from 1.32 rad to the disk midplane. We simulate a 0.05 M$_\odot$ disk around a 1 M$_\odot$ star. We include a 318 M$_\oplus$ giant planet at a fixed separation of 75 au using a mass taper over the first 100 orbits. The disk aspect ratio at the planet separation was set to 0.05.  The disk is simulated for 1000 orbits, and the resulting structure is used to determine the dust temperature in post-processing, following \citet{alarcon_thermal_2024}. The resulting gas density and velocity structure are shown in Figure \ref{fig:disk_structure}, with the velocity shown relative to the planet, $\tilde{v}$, with $\tilde{v}_r < 0$ indicating gas flow towards the planet. Near the midplane, gas is primarily flowing away from the planet along the spiral arms. Gas accretion tends to be onto the poles of the planet, as is expected from meridional flow patterns \citep{kley_three-dimensional_2001, morbidelli_meridional_2014, szulagyi_accretion_2014}. Within the pressure bump near 100 au, gas advection tends to be vertically away from the midplane.

The temperature within the disk is determined using RADMC-3D \citep{dullemond_radmc-3d_2012} including radiation from the central star, emission from the embedded planet, and viscous heating \citep[see][]{alarcon_thermal_2024}. For the temperature structure, we assume a constant dust-to-gas mass ratio of 0.01, with the solid mass entirely in small dust. The outer disk is characterized by passive heating, with the upper layers of the disk being hotter than the regions close to the midplane \citep{chiang_spectral_1997}.
In the gap, a combination of accretion heating from the planet, viscous heating at the midplane, and asymmetric dust scattering results in increased temperatures near the midplane, as shown via the isothermal contour in Figure \ref{fig:disk_structure}.
% Near the planet, accretion heating raises the local temperature, increasing the dust and gas temperature at the midplane in the gap, as can be seen via the isothermal contour in Figure \ref{fig:disk_structure}. 

As an example of a volatile species, we track the sublimation of CO from the grains at this location in the disk. CO is highly volatile in protoplanetary disks; the temperature at which it begins to appreciably come off the grains depends on the CO binding energy and surface area of grains available.
The binding energy of CO is uncertain and depends on the ice mixture on the mantle of the grain, with literature values ranging from 855~K \citep{mcelroy_umist_2013} up to 1500~K \citep{fayolle_n2_2016}, suggesting freeze-out temperatures of $\sim$20 - 30~K.
Throughout this work, we use a conservative assumption for the CO sublimation temperature of 30~K, corresponding to a midplane snowline location of 45~au, interior to the planet location of 75~au modeled here.
The location of the CO snowline is shown by the red dashed line in Figure \ref{fig:disk_structure}.
While this temperature is an approximation, the results presented here do not depend the exact temperature, and we discuss differing heights of the CO sublimation surface later.
Although, heating from the embedded planet increases the midplane temperature above the CO sublimation temperature near the edge of the gap, the main region of the pressure bump that is above 30~K is above $Z/R\approx0.1$, slightly above 1 scale height at this location of the disk.

Small solids in the disk, which serve as the feedstock of planetesimals and planets, can be grouped via their aerodynamic properties and interactions with the gas. These aerodynamic properties are defined via the Stokes number, St, which is the dimensionless ratio of the stopping time to the orbital period of the solid. Throughout this work, we refer to two main populations of solids, the small dust and larger pebbles \citep[as in, e.g.][]{birnstiel_simple_2012}. Small dust (St~$<$~$10^{-3}$) is defined as those solids which are primarily coupled with the gas, and whose dynamics are strongly affected by gas advection and turbulence. The dynamics of larger pebbles ($10^{-3}$~$<$~St~$\lesssim$~1) are mainly driven via gas drag-induced radial drift \citep{weidenschilling_aerodynamics_1977}. Pebbles also experience significant settling, with typical scale heights much less than that of the gas, in contrast with the smaller dust which has a comparable scale height to the gas \citep{youdin_particle_2007}.

Once the physical conditions within the disk are known, we track the location and temperature of pebble and dust tracer particles of constant size using the 3D Monte-Carlo particle tracking technique described in \citet{van_clepper_three-dimensional_2025}. Particles are started at the midplane exterior to the gap opened by the giant planet, here at about 100 au. We examine particle sizes ranging from submicron, St < $10^{-5}$, up to 1 cm in size, St $\approx$ 1. Because the solids are integrated using a constant particle size, their Stokes numbers vary throughout the integration, with the Stokes number inversely proportional to the gas density. As a result, particles' Stokes numbers decrease in the pressure bump, where gas densities are at a local maximum, and increase in the gap and away from the midplane, where gas densities are lower. Where Stokes numbers are given here, it is the Stokes number for solids at the midplane at 100 au as a means to compare between the 3D constant size integration and 1D DustPy simulations, described further below.
Particles of all sizes are integrated forward in time including random diffusion for 1 Myr, with their locations recorded every 10 years ($\sim$ 0.01 orbital period). At each time, the temperature of the particles are interpolated from the RADMC-3D outputs. While the RADMC-3D simulations only consider the small, well-coupled dust to determine the dust temperature, we assume that all dust and gas are in thermal equilibrium; that is, the temperature of all sizes of dust and the gas is the same at a given location of the disk.

\begin{figure}
    \centering
    \includegraphics[width=\linewidth]{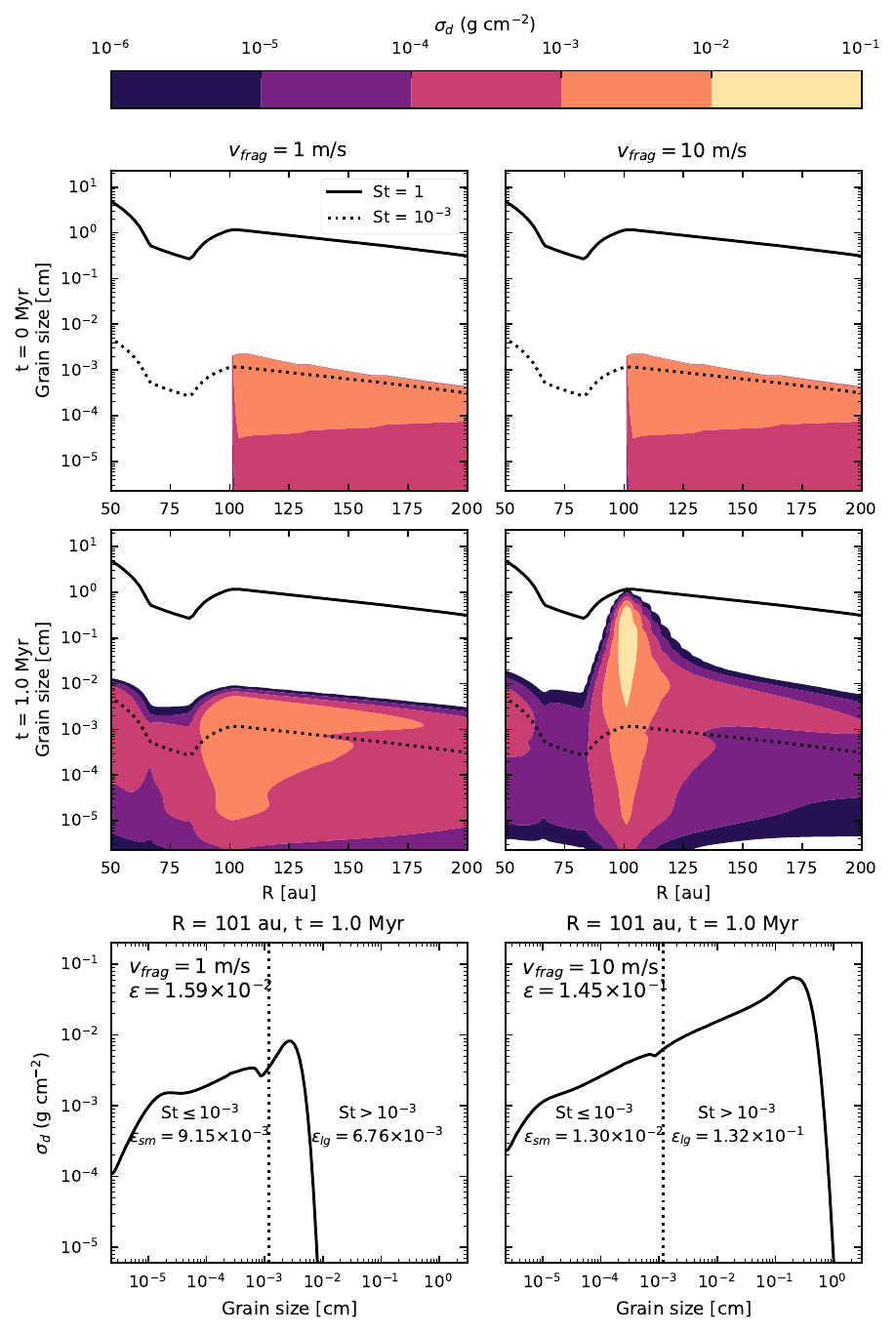}
    \caption{Dust distribution as a function of size and radius from DustPy simulations showing the initial conditions of the dust size distribution (top row) and after 1 Myr of evolution (center row) for a disk containing a Jupiter mass planet located at 75 au. We examine disks with particle fragmentation velocities of 1~m~s$^{-1}$ (left column) and 10~m~s$^{-1}$ (right column). The grain size distribution in the pressure bump at 101 au is shown in the bottom row. Regardless of dust fragmentation velocity, we see an enhancement of small dust less than St = $10^{-3}$ (dotted line) in the pressure bump. The total dust-to-gas mass ratio, $\varepsilon$, is labeled in each plot, where the initial dust-to-gas is $\varepsilon$ = 0.01 in the outer disk. The dust-to-gas mass ratios for grain with St $\leq 10^{-3}$, $\varepsilon_{sm}$, and St $> 10^{-3}$, $\varepsilon_{lg}$ are also shown. For the case with larger fragmentation velocity, while the total dust mass is higher and grains reach much larger sizes, the total mass of dust with St $\leq 10^{-3}$ is about 0.01 in both cases.}
    \label{fig:dustpy_results}
\end{figure}

While particles in our simulations remain at fixed sizes, we also use DustPy simulations \citep{stammler_dustpy_2022} to model the dust size evolution for the disk. We match the DustPy surface density to the FARGO3D simulation including a Jupiter mass planet at 75 au. The disk is initialized with a MRN-like dust size distribution \citep{mathis_size_1977} in the outer disk with a dust-to-gas mass ratio of 0.01. As the fragmentation velocity of dust is uncertain, we do two simulations, one with a fragmentation velocity of 1~m~s$^{-1}$ and one with 10~m~s$^{-1}$. Both simulations assume a constant $\alpha$ viscosity parameter of $\alpha = 10^{-3}$ as in our FARGO3D runs. The initial conditions and dust size distribution after 1 Myr of evolution for both simulations are shown in Figure \ref{fig:dustpy_results}. The location of the pressure bump in the DustPy simulations is centered at 101 au, agreeing with the full 3D hydrodynamic simulations.

\section{The DRIVE mechanism}
\label{sec:results}

\begin{figure}
    \centering
    \includegraphics[width=\linewidth]{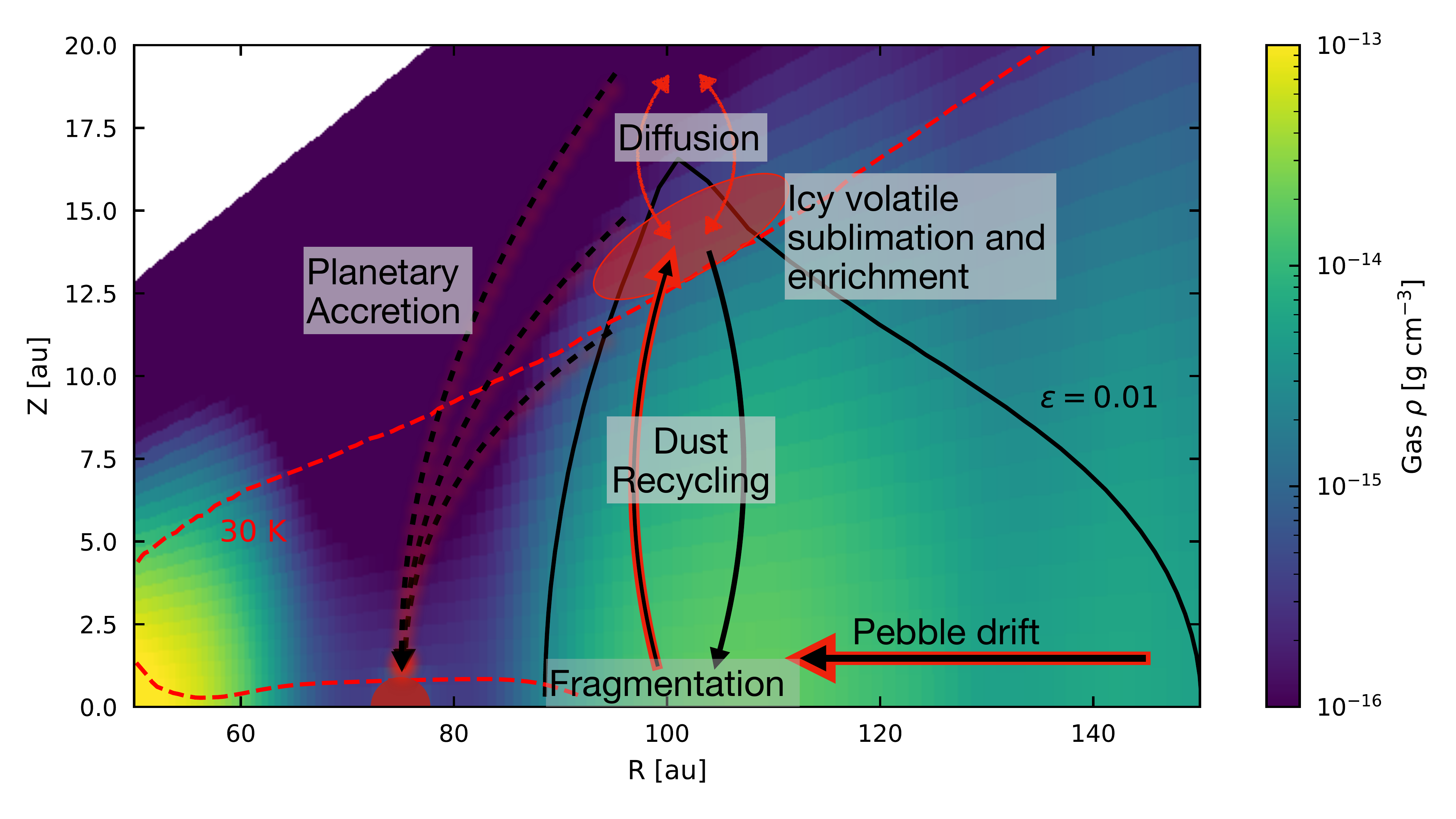}
    \caption{Cartoon schematic of the DRIVE effect enriching the atmosphere of a giant planet with volatiles. Larger pebbles trap volatiles from the outer disk and transport them to the pressure bump created by the giant planet via radial drift. Here, fragmentation creates fine dust, which can be lofted away from the midplane, leading to sublimation of ice mantles on the dust and enriching the gas above the snow surface with volatiles. This volatile rich gas can then be accreted onto the planet, resulting in a volatile enriched atmosphere. The dust, due to the lower gas density at the surface of the disk, decouples from the gas and can settle back to the midplane as a bare grain, fractionating the volatile elements from the more refractory solids. Here, the color shows the azimuthally averaged gas density from our FARGO3D simulations. The solid black contour outlines the region of the disk where total dust-to-gas mass ratio is equal to 0.01 in our DustPy simulations, with higher dust-densities below this contour. The transport of icy pebbles is represented by the black and red solid arrows, while black arrows represent bare grains and red arrows represent CO gas. Meridional flows onto the planet including entrained CO gas are illustrated by the red and black dashed arrows. The approximate location of the CO snowline at 30~K is also included.}
    \label{fig:schematic}
\end{figure}

Combining these techniques, we examine the stirring of small dust in the pressure bump created by a giant planet. We find that a natural consequence of dust growth is the concentration of solids in the dust trap and stirring of small dust to higher and warmer regions of the disk, enriching the gas accreted onto the planet in volatiles.
Figure \ref{fig:schematic} shows a cartoon illustration of this process highlighting the essential ingredients: pebble drift, fragmentation, meridional circulation, and volatile sublimation.
We refer to this process as ``Dust Recycling and Icy Volatile Enrichment'' (DRIVE), whereby small dust grains undergo a cycle of growth and fragmentation, transporting volatile ices above the snow surface and sublimating them into the gas phase. The gas, now enriched in volatiles sublimated from the grains, can then be accreted onto the planet and incorporated into the giant planet atmosphere.
Thus, we do not require giant planets to grow interior of the midplane snowline to inherit volatile gas into their atmosphere. This provides a novel way to enrich the atmospheres of cold giant planet after they have reached the pebble isolation mass.

\subsection{Solid concentration and fragmentation}
\label{sec:dustpy-results}

Once they reach the pebble isolation mass, giant planets create pressure maxima external to their orbit \citep{rafikov_planet_2002}, where radial drift is halted and solids can concentrate \citep{weidenschilling_aerodynamics_1977, lambrechts_separating_2014, dullemond_disk_2018}.
In these regions, grains grow to a certain maximum size before relative velocities exceed fragmentation velocities, breaking apart larger pebbles into small dust grains \citep{birnstiel_dust_2011, birnstiel_simple_2012}. Indeed, numerical simulations have shown that fragmentation is a common outcome in the pressure bump exterior to a giant planet \citep{drazkowska_including_2019, stammler_leaky_2023, eriksson_particle_2024}.

Our DustPy simulations show that regardless of fragmentation velocities assumed, the pressure bump exterior to the planet becomes enriched in both large pebbles and small dust. Because these dust grains in the bump are supplied by inward drift of ice-rich pebbles, the increase in dust mass also corresponds with an increase in total CO in the bump. Figure \ref{fig:dustpy_results} shows the grain size distribution in the pressure bump, where both the dust-to-gas mass ratio, $\varepsilon$, and maximum grain size reach peak values. We also show the dust-to-gas mass ratios for small and large grains, $\varepsilon_{sm}$ and $\varepsilon_{lg}$ respectively, where we define small grains as dust (St$\leq 10^{-3}$) and large grains as pebbles (St$> 10^{-3}$).
In both the model with low fragmentation velocity and high fragmentation velocity, the total dust-to-gas mass ratio is increased above the initial value of $10^{-2}$, with $\varepsilon = 1.59\times10^{-2}$ and $1.45 \times 10^{-1}$ respectively after 1 Myr.
Despite differences in the total dust-to-gas ratios, the value of $\varepsilon_{sm}$ is comparable in both cases, $\varepsilon_{sm} = 0.915\times10^{-2}$ and $1.3\times10^{-2}$ for $v_{frag} = 1$~m s$^{-1}$ and 10~m s$^{-1}$ respectively.
Regardless of fragmentation velocity, we see an enhancement not only in the total dust-to-gas mass ratio, but also an enhancement of the small, well-coupled dust with St $\leq 10^{-3}$, approximately equal to the initial value of 0.01 in the pressure bump exterior to the planet. This enhancement in small dust is key to the delivery of CO ice to the gas, as the small dust is more mobile in the pressure bump and can be lofted to regions in the disk above the CO sublimation temperature, sublimating CO from the ice phase into the gas.

\subsection{Dust recycling and ice sublimation}
\label{sec:partrace-results}

\begin{figure}
    \centering
    \includegraphics[width=\linewidth]{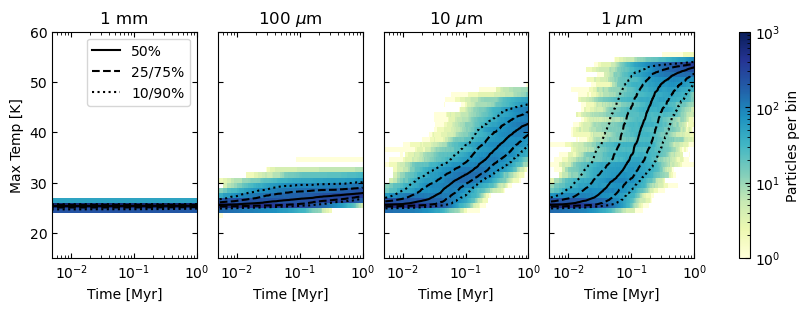}
    \caption{The maximum temperature reached by particles of different sizes, ranging from 1~mm to  1~{\textmu}m (right column) for particles starting at the midplane at 100~au. The colormap shows the number of particles out of 1000 total that have reached a given temperature as a function of time. Also indicated in each plot by a solid line is the 50\ts{th} percentile for maximum temperature reached as a function of time. The dashed lines show the 25\ts{th} and 75\ts{th} percentiles, and dotted line show 10\ts{th} and 90\ts{th} percentiles. While large particles are not heated much above the midplane temperature, solids below 10~{\textmu}m in size (St $\lesssim 10^{-3}$ at the midplane experience heating. All of these dust particles experience temperatures above 30~K by 1~Myr.}
    \label{fig:partmaxtemps}
\end{figure}

Next, we examine the thermal histories of solids of different sizes immediately exterior to the planet.
The larger pebbles, here ranging from 100~{\textmu}m to 1~cm, tend to stay near the cold midplane, while grains smaller than 10~{\textmu}m can be lofted to higher elevations and reach higher temperatures.
Figure \ref{fig:partmaxtemps} shows the maximum temperature reached by 1000 particles of different sizes over time. 
Dust with sizes~$\leq 10$~{\textmu}m (St $\leq 10^{-3}$) experience temperatures much higher than the midplane temperature of 21~K. After 0.1 Myr, 75\% of micron-sized dust has reached temperatures well above the CO sublimation temperature of 30~K with 100\% of the small dust reaching temperatures above 30~K by 1 Myr. This trend is similar for 10~{\textmu}m, while the larger 100~{\textmu}m show only about 10\% of grains reaching above 30~K, with no millimeter or larger grains reaching the CO sublimation temperature. These larger grains tend to experience some heating from the luminous planetary embryo, but here it is insufficient for CO sublimation, and we focus on the heating of small grains at the surface of the disk.

\begin{figure}
    \centering
    \includegraphics[width=\linewidth]{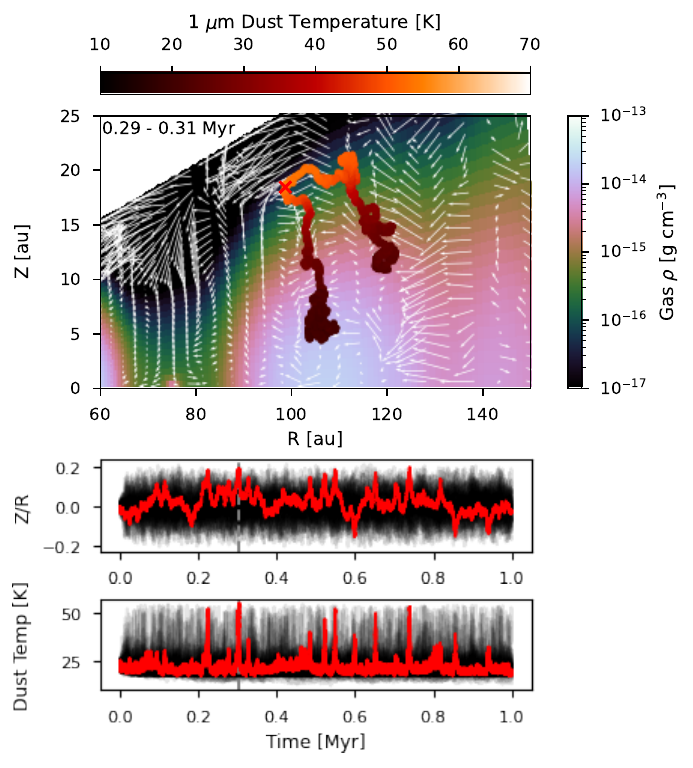}
    \caption{(Top) Particle trajectory of the tracer particle reaching the maximum temperature, with the hottest point labeled with a red dot. The background color and arrows show the azimuthally averaged gas density and advection. This particle is lofted to high altitude following the meridional gas circulation in the disk, reaching temperatures up to 55~K. The particle's temperature along its trajectory is shown by the color of line. The particle temperature is shown over 20 kyr to demonstrate the particle's trajectory before and after reaching the hottest point at 0.3 Myr. This particle's height above the midplane (middle) and temperature (bottom) are also shown as a function of time over the full 1 Myr integrated by the red lines. Examples of other particles are also shown in black, indicating the highlighted particle's trajectory and temperature are typical of other similarly sized particles.}
    \label{fig:temp_overtime}
\end{figure}

The locations and temperatures over time for one hundred 1~{\textmu}m sized examples particles are shown in Figure \ref{fig:temp_overtime}. Although these small dust grains tend to experience the highest degree of heating, they spend the majority of the time near the midplane and low temperatures. When these solids are heated, it is typically during brief excursions to upper layers of the disk, before returning to the cold midplane, as has been found in studies of disks not containing a planet \citep[e.g.][]{ciesla_organic_2012, bergner_ice_2021, flores-rivera_uv-processing_2024}.

This migration of small grains to the disk surface is enhanced via the meridional flows created by the planet.
3D hydrodynamic simulations have shown that planet-induced pressure bumps can significantly increase the scale height of small dust \citep{bi_puffed-up_2021, szulagyi_meridional_2022}. In Figure \ref{fig:temp_overtime}, we also show the trajectory of the particle that experiences the highest degree of heating near the time when it reaches this temperature. The planet creates meridional flow patterns that can entrain small particles, lifting them near the surface of the disk. Here, where the dust grains are heated, they sublimate CO ice to the gas phase, increasing the CO gas abundance above the snow surface.
In Figure \ref{fig:temp_overtime}, the highlighted particle's position and temperature is shown over only 20 kyr, yet the dust particle is transported vertically over 2 gas scale heights. While the timescale for vertical transport via turbulence in the disk is $\tau_{turb} = (\alpha\Omega)^{-1} \approx$ 200 kyr for $\alpha=10^{-3}$, dust particles simulated here tend to be transported on much shorter timescales, indicating gas advection as the dominant transport mechanism.

While these grains initially follow the gas advection away from the midplane, they do not tend to be accreted by the planet. This is because as the small grains are lofted, the gas density decreases, resulting in an increase to the particle's Stokes number. Thus, while experiencing heating at the surface of the disk, the grains decouple from the gas flow and settle back to the midplane, where -- now depleted in CO ice -- they may coagulate into larger pebbles \citep{krijt_dust_2016, misener_tracking_2019}. Any gas phase CO that is not accreted onto the planet may diffuse back below the snow surface and freezeout again as ice. Even if some amount of CO vapor returns to the grain, that grain remains in the pressure bump, where fragmentation will resupply the fine dust, which can again transport that CO ice back to the disk surface, repeating the cycle of grain lofting and volatile sublimation. As a result, a single micron sized grain may transport CO ice to the surface of the disk multiple times over the disk lifetime.

\subsection{Giant planet atmosphere enrichment}
\label{sec:volatile-funneling}

We have shown that embedded giant planets are expected to both create dusty pile-ups exterior to their orbit and that dust grains in this pile-up should experience heating when they advect to the upper regions of the disk.
These small grains are resupplied via the fragmentation of inward drifting, volatile-rich pebbles from the outer disk.
As a result, volatiles are carried to the surface regions of the disk, where they can sublimate and be accreted onto the planet, even in the case where the planet is located outside of the midplane snowline.

To determine the magnitude of enhancement we expect above the snow-surface, we assume a steady-state dust distribution based on our DustPy simulations.
We assume that this disk structure has evolved from an initially homogeneous disk, with constant dust-to-gas and CO abundance given by $\varepsilon_0$ and $\XCOi$ respectively, where the dust-to-gas is given by $\varepsilon_0 = \rho_{dust}/\rho_{gas} = 0.01$. Outside of the CO snowline, CO is entirely frozen onto grain surfaces such that the CO-to-grain mass ratio remains constant throughout the disk evolution.
Additionally, we assume that grains heated above 30~K are able to fully sublimate CO from the ice-phase to the gas phase, such that the CO gas abundance, $\XCO$, is proportional to the mass of grains heated above 30~K.
Finally, we assume that in this steady-state approximation CO is able to diffuse vertically, such that $\XCO$ is constant everywhere above the snow surface.

\begin{figure*}
    \centering
    \includegraphics[width=0.9\linewidth]{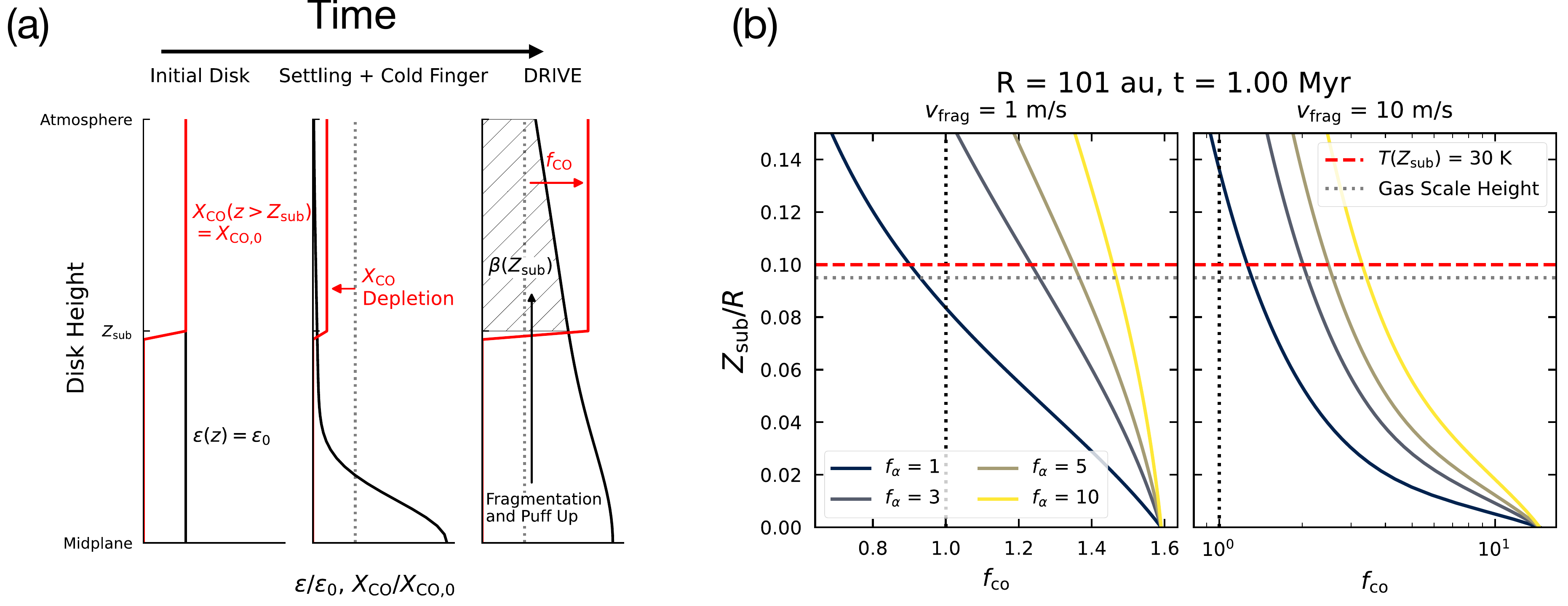}
    \caption{\emph{(a):} Schematic diagram illustrating the vertical stirring of dust and expected CO enrichment. Initially, the vertical column has a uniform dust-to-gas mass ratio, $\varepsilon$, shown in black, and gas-phase CO abundance $\XCO = \XCOi$, shown in red, above the sublimation height, $Z_\text{sub}$ (top). Typically, dust growth and settling concentrated dust mass near the midplane, with CO gas depleted as a result of the cold finger effect (middle). In the DRIVE Mechanism, dust is puffed vertically from planetary stirring carrying CO ice above the sublimation height. CO ice is then sublimated into the gas phase, with the abundance of CO proportional to the integrated dust-to-gas mass ratio above $Z_\text{sub}$, indicated by the hatched region.  {\color{blue}} \emph{(b):} Expected CO abundance relative to the initial abundance, $f_\text{CO}$, as a function of $Z_\text{sub}$ and $f_\alpha$. $f_\text{CO}$ is calculated from the DustPy simulation after 1 Myr of evolution in the pressure bump for fragmentation velocities of 1~m~s$^{-1}$ (left) and 10~m~s$^{-1}$ (right). The CO sublimation surface where dust temperatures reach 30~K is shown by the red dashed line, while the gas scale height is shown by the gray dotted line. The dust scale-height is increased using a diffusivity of $f_\alpha \alpha$, accounting for vertical stirring as a result of meridional flows. We find that for $Z_\text{sub}/R < 0.1$, dust densities (and thus CO abundances) are near or above expected solar metallicities in all cases.}
    \label{fig:beta_stirring}
\end{figure*}

This process is outlined in Figure \ref{fig:beta_stirring}, showing a schematic of the dust-to-gas and CO abundance in a column of the disk over time. Initially, the dust-to-gas mass ratio is constant throughout the column, and CO is only present as gas above $Z_\text{sub}$, where $Z_\text{sub}$ is the height above the midplane where ice sublimation occurs. For CO sublimation at $T=30$~K, this is located at $Z/R\approx0.1$ at 100 au, but $Z_\text{sub}$ will depend on the radial location in the disk and sublimation temperature for the specific volatile species of interest. If the dust in the column is puffed up, as we expect from giant planet induced stirring, then these grains will carry CO ice above the sublimation surface, increasing the abundance of CO there.

Under these assumptions, the total abundance of CO gas above the snow surface is given by
\begin{equation}
\label{eq:co_abun}
    \XCO = 
    \frac{
        \beta(Z_\text{sub})
    }{
        \varepsilon_0
    } \XCOi = 
    f_\text{CO} \XCOi,
\end{equation}
where $\beta(Z_\text{sub})$ is the integrated dust-to-gas ratio above $Z_\text{sub}$. The enhancement in CO above the initial abundance is given by $f_\text{CO} = \beta(Z_\text{sub})/\epsilon_0$, where $\beta(Z_\text{sub})$ is given by the equation
\begin{equation}
    \beta(Z_{sub}) =
    \frac{
        \int_{Z_\text{sub}}^\infty
            \int_{a_0}^{a_\text{max}}
                \eta(a,z)
            \text{ d}a
        \text{ d}z
    }{
        \int_{Z_\text{sub}}^\infty
            \rho(z)
        \text{ d}z
    }.
\end{equation}
Here, $\rho(z)$ is the gas density and $\eta(a,z)$ is defined such that $\int_{a_0}^{a_{max}} \eta(a,z)\text{ d}a = \rho_\text{dust}(z)$ gives the total volume density of dust at height $z$, with $a_0$ and $a_\text{max}$ being the smallest and largest grains present. 
By definition, $\beta(0)$ (where $Z_\text{sub}/R=0$ corresponds to a location interior to the midplane volatile snowline) is equivalent to the column integrated dust-to-gas mass ratio. \added{This corresponds to the total $\epsilon$ from our DustPy simulations (See Section \ref{sec:dustpy-results} and Figure \ref{fig:dustpy_results})}. Inside of the midplane snowline, all of the volatile ice is expected to be sublimated. \added{At this location where $Z_\text{sub}/R=0$, $f_\text{CO}$ is precisely the enhancement in dust-to-gas relative to the initial $\epsilon_0$ integrated vertically. Exterior to the midplane snowline ($Z_\text{sub}/R>0$), some amount of volatiles remain in the ice phase near the midplane, with the abundance of sublimated volatiles proportional to the dust mass above the sublimation temperature.} 

\added{Here, we assume a gas with a typical ISM dust-to-gas mass ratio of $\epsilon_0=0.01$ corresponds with a solar metallicity and that the total metallicity of the gas is proportional to the dust-to-gas mass ratio above the sublimation temperature, as the majority of molecules other than hydrogen and helium are frozen as ice onto the grains. As such, $\beta(Z_\text{sub}) > \epsilon_0$ represent cases in which the gas is enhanced in CO, corresponding to a super-solar metallicity, and $f_\text{CO}>1$.}

Because DustPy is a 1D simulation, we assume a vertical distribution of dust and grains from the surface density and gas scale height. The gas scale height, $h = c_s\Omega$ (where $\Omega$ is the local Keplerian frequency and $c_s$ is the sound speed), is assumed from a \citet{chiang_spectral_1997} like midplane temperature profile. The dust scale height, $h_a$, is written as in \citet{youdin_particle_2007}:
\begin{equation}
    h_a = \sqrt{\frac{\delta}{\delta + \text{St}_a}}h,
\end{equation}
where $\delta$ is the vertical diffusivity of the dust and St$_a$ is the Stokes number of a solid of size $a$. We define $\delta = f_\alpha\alpha$, where $f_\alpha$ is a dimensionless enhancement parameter to the vertical turbulence, to account for an increase in the vertical scale height of dust as a result of stirring via meridional flows exterior to the planet \citep{bi_puffed-up_2021, szulagyi_meridional_2022}. While this value of $f_\alpha$ is uncertain, we test a variety of values ranging from $1\leq f_\alpha \leq 10$.

The enhancement in CO relative to the initial abundance, $f_\text{CO}$, is plotted as a function of $Z_\text{sub}/R$ at $R$ = 101 au, the location of the pressure bump and dust trap for our DustPy simulations, in Figure \ref{fig:beta_stirring}. Here, we see that higher fragmentation velocities and stirring both tend to increase the mass of dust above the sublimation surface, and thus the expected abundance of gaseous CO. For $Z_\text{sub}/R = 0.1$, as is the case for CO at 101 au, CO abundances directly exterior to the planet range from slightly sub-solar up to a maximum of 3$\times$ solar abundance if $f_\alpha = 10$. While disk turbulence is difficult to constrain, observations of disks containing ring and gap structures have shown that some rings may be ``puffier'' than others, with vertical turbulence up to $\alpha \approx 10^{-2}$ \citep{doi_estimate_2021, villenave_turbulence_2025}, corresponding to $f_\alpha = 10$ in our model. We note that 30~K assumed here is likely an upper limit for the temperature of CO sublimation. If CO begins to sublimate off of the grain at lower temperatures, this will correspond with lower values of $Z_\text{sub}$ and an increase of $\beta(Z_\text{sub})$. Thus a lower sublimation temperature -- and lower sublimation height in the disk -- will increase the expected volatile enrichment onto the planet. A sublimation temperature of 25~K, for example, corresponds to $Z_\text{sub}/R\approx0.09$. For this lower sublimation temperature, and thus lower sublimation height, CO abundances range from solar up to 4$\times$ solar.

While low fragmentation velocities tend to create more small dust, the concentration of total solids in the pressure bump remains relatively low compared with the higher fragmentation velocity model (See Section \ref{sec:dustpy-results} and Figure \ref{fig:dustpy_results}). For the high fragmentation velocity case, pebble growth and inward drift concentrates solids in the bump, 
\added{leading to a higher total dust-to-gas mass ratio at this location. For the case with a fragmentation velocity of 10~m~s$^{-1}$, this leads to a total enhancement of 14.5$\times$ the initial $\epsilon_0$, shown as $f_\text{CO}(Z_\text{sub}/R=0)$ in Figure \ref{fig:beta_stirring}.
While higher fragmentation velocities lead to larger pebbles, the greater concentration of solids in the pressure bump also increases the mass of small dust, with the ``tail'' of the mass distribution towards smaller grains resulting in significant dust mass above 30~K even for the case with no additional vertical stirring ($f_\alpha = 1$).}

The maximum enhancement to dust in the upper disk, and thus enhancement in CO gas, would occur from efficient inward drift of pebbles from the outer disk followed by fragmentation in the bump. We note that fragmentation in the pressure bump exterior to a giant planet is expected to be enhanced, due to increased relative velocities in the spiral arms \citep{drazkowska_including_2019, eriksson_particle_2024}.
Additional vertical stirring as a result of planetary meridional flows, as expected from 3D simulations \citep{bi_puffed-up_2021, szulagyi_meridional_2022}, will act to further increase the delivery of volatile ice above the CO snow surface.

The ingredients for the DRIVE mechanism are expected to be present regardless of where the planet forms.  The detailed implications for the composition, including the C/O, of the planet will depend on its formation location relative to the various snow lines that would exist throughout the disk.  For example,  CO$_2$ may also sublimate off the lofted grains considered here, but due to its higher binding energy, will only do so for those grains that reach temperatures in excess of $\sim$50~K.  This would lead to a lower enhancement in CO$_2$ than CO because of the former’s higher snow surface.   Were the planet forming just outside the midplane CO$_2$ snow line, however, the planet may receive an enrichment in CO$_2$, while the CO would be present at roughly solar abundance as its snow line is located much further out in the disk and drifting pebbles would be absent of CO ice.  If appreciable CO$_2$ is accreted along with CO, this would push the C/O of the planetary atmosphere below unity, possibly explaining C/O ratios near solar in wide-separation giant planet atmospheres \citep{bergin_co_2024, nasedkin_four---kind_2024}.

Here we highlight the importance of the 2D snow \emph{surface}, rather than focus on the midplane snowline. Previous studies have highlighted the ``cold finger'' effect, where dust growth and settling can deplete volatiles from the surface layer of the disk, enriching the midplane \citep{meijerink_radiative_2009, krijt_tracing_2016, van_clepper_chemical_2022}. In the presence of a giant planet, however, the dust pile-up will lead to frequent collisions and fragmentation, replenishing the supply of small grains exterior to the planet. These small grains cycle between the midplane and upper regions of the disk on 10 kyr timescales (less than the diffusive mixing time), carrying their volatile ice mantle with them.
Here, grains sublimate their volatiles into the gas phase, creating a volatile rich gas environment above the snow surface. Given the giant planet accretes gas from above one scale height, this volatile rich gas will be readily accreted into the atmosphere of the giant planet.

\subsection{Caveats to the model}

The magnitude of the volatile delivery onto the planet will depend in detail on the production of fine dust from pebbles and the heating of that dust, which itself depends on the interplay between grain growth, fragmentation, transport, and heating in 3D. Modeling these processes together, however, is a computationally expensive task and will depend on many factors, including the disk mass, viscosity, accretion rate, and pebble flux. Here, we examine some of the assumptions and limitations of the modeling work presented here.

First, we assume that sublimated volatiles above the snow surface are accreted onto the planet. Hydrodynamic models of cooling giant planet atmospheres show that meridional flows are primarily recycling flows, with the majority of gas returning to the disk \citep{kurokawa_suppression_2018, bailey_growing_2024}. Nonetheless, any gas that is accreted onto the planet will come from the surface regions of the disk, which will be enhanced in volatiles as a result of the DRIVE mechanism. While only a small fraction of disk gas may ultimately be incorporated into the giant planet atmosphere, any disk gas that is incorporated will have an enhanced metallicity. Additionally, we emphasize the \emph{recycling} nature of the DRIVE effect. In the case where volatile enriched gas is not accreted onto the planet or adsorbs onto another grain, that grain may be then cycle to the upper disk, desorbing volatiles into the gas phase again. While the overall metallicity enhancement onto the planet may require a detailed understanding of the gas accretion onto giant planets, the DRIVE mechanism presented here predicts that disk gas available for accretion should be enriched in volatiles above the snow surface by a factor of up to 3$\times$ solar.

We also examine our dust population assumptions from the 1D DustPy model presented in Section \ref{sec:dustpy-results}. In 2D, for example, when small solids are lofted they will collide with other small solids, resulting in growth via sweep-up. \citet{krijt_dust_2016} studied dust growth and settling in a 1D vertical slice of an axisymmetric disk in the absence of a planet. They showed that even for high dust-to-gas ratios up to 0.1, as would be expected in the pressure bump exterior to the planet, small dust grains can still reach heights up to the gas scale height before growing and settling back to the midplane. \citet{misener_tracking_2019} expanded this to a 2D disk ($R-Z$) including growth, fragmentation, settling, and drift, and showed similar results, with small grains typically reaching 1-2 gas scale heights in the disk shortly after a collision before growing and settling again. Here, though, we again emphasize the role of the meridional flows, which will serve to further increase the dust scale height, as shown here and in other work \citep{bi_puffed-up_2021, szulagyi_meridional_2022}. Thus, we expect even in the case of dust-sweep up at altitude, small grains should be able to deliver volatiles to the warm atmosphere of the disk.

These 3D models of dust transport around giant planet not only show that dust is puffed up at the edges, but also that small dust grains can be accreted onto the planet \citep{szulagyi_meridional_2022, van_clepper_three-dimensional_2025} and that a dust envelope (or disk) may be expected around the embedded planet \citep{bi_puffed-up_2021, krapp_3d_2022}. However, in the surface regions of the disk and in the gap near the planet, radiative transfer models predict that both temperatures and UV irradiation should be higher than elsewhere in the disk. \citep{alarcon_chemical_2020, alarcon_thermal_2024}. Thus, except for the in the pressure bump where the grain may be recycled back into the DRIVE mechanism, the recondensation of sublimated volatiles onto grains is unlikely.

Next, we consider how this effect might operate at other snowlines throughout the disk. For grains to be heated as they are lofted in the disk, this requires the gas temperature to be lower at the midplane and higher at the surface of the disk. While this is true for disks where heating is primarily a result of irradiation, viscous heating at the midplane can also provide a source of heat. Viscous heating depends on the mass accretion rate and disk mass \citep{calvet_flat_1994}, and as a result is more dominant in the inner disk where midplane densities are highest. As such, DRIVE should be most apparent in the outer disk, where the surface of the disk is hotter than the midplane. ``Cold'' snow surfaces, such as the CO and CO$_2$ sublimation fronts, are thus most likely to undergo this process of grain lofting and sublimation, while ``hotter'' snowlines, like the H$_2$O snowline or ``soot line'' \citep[location of refractory carbon sublimation, ][]{kress_soot_2010, lee_fire_2010, li_carbon_2021}, are more likely to have colder disk surfaces, and thus may not be as susceptible to thermal desorption at the surface of the disk. Enhanced UV radiation at the surface of the disk, however, may drive grain chemistry including desorption in these regions \citep{ligterink_rapid_2024}. UV-driven photo desorption may lead to sublimation of even water ice, further increasing the metallicity and decreasing C/O of the disk gas.

Finally, as dust grains are the primary transport mechanism for volatiles in the DRIVE model, grain surface chemistry will play an important role setting the gas-phase volatile abundances. Irradiation, for example, may dissociate CO into atomic C and O, resulting in the production of secondary ice species with different binding energies \citep[see, e.g.,][]{bergner_ice_2021, ligterink_rapid_2024, cridland_gas_2025}. If the resulting primary C- and O-bearing have drastically different binding energies from one another (or from CO) then this may effect the resulting metallicity and C/O of volatiles sublimated from the grains and ultimately the planetary atmosphere. This will, however, depend on the relative timescales involved between grain surface and gas phase chemistry and optical depth of the dust, and should be studied in more detail using dynamic chemical models.

\section{Comparison with other mechanisms}
\label{sec:comparisons}

We present the DRIVE mechanism as an additional means by which giant planet atmospheres may become enriched in volatile species.
While the different mechanisms of giant planet atmosphere enrichment are not mutually exclusive from one another, each may be dominant in different disk conditions and leave distinct traces of that formation pathway on the formed planet.
In this section, we briefly summarize some such models and differences with our model presented here.

\subsection{Pebble drift across snowlines}

The process here is similar to the enrichment of volatiles at snowlines as a result of pebble drift \citep{oberg_excess_2016, booth_chemical_2017, zhang_excess_2020},
although these models require the planet form interior to the snowline, while the DRIVE model presented here allows for planet to incorporate these volatile even when forming outside of the midplane snowline.
Enrichment via pebble drift across snowlines requires large pebbles to drift past midplane snowlines, enhancing the abundance of a given volatile species and imprinting an atmospheric C/O based on the sublimated ices. Here, we enhance the gas-phase with volatile species via the vertical lofting of small grains above the snow surface.
Additionally, the pebble drift model of enhancement would predict that CO should be enhanced in the inner disk as CO-rich pebbles drift across the CO snowline.
Disk observations, however, have shown that CO is depleted rather than enhanced interior to the CO snowline \citep{zhang_systematic_2019}.

\subsection{Planetesimal Accretion}
Another way to enhance the metallicity of the atmosphere of giant planets is via accretion and ablation of planetesimals. If these planetesimals originate in the outer disk, then they may be volatile rich \citep{atreya_comparison_1999}.
Jupiter, for example, may have formed in the outer disk, beyond the N$_2$ snowline, accreting C, N, and O from icy planetesimals \citep{bosman_jupiter_2019, oberg_jupiters_2019}, before migrating to its current day location.
This inward migration may further act to shepherd planetesimals onto the planet, enriching the envelope of closer in exoplanets \citep{shibata_origin_2020, shibata_origin_2022} and Jupiter \citep{shibata_enrichment_2022} if it migrated in from the outer disk. If these planetesimals retain all volatiles during their accretion, then this may lead to decreased C/O of the planetary atmosphere closer to solar values.
This assumes there are ample planetesimals in the disk to be accreted, however dynamical models show scattering, rather than accretion, may dominate interactions between planets and planetesimals \citep{eriksson_low_2022}.

An important aspect of the DRIVE model is the separation of volatiles from their refractory carriers, resulting in preferential accretion of volatiles over solid dust grains. This refractory-to-volatile ratio in the atmospheres of planets is expected to differ depending on the mechanism of enrichment \citep{lothringer_new_2021}. Thus this refractory-to-volatile ratio may help to differentiate between planetesimal accretion, pebble accretion, and the DRIVE mechanism .
The DRIVE model presented here provides a novel way to transport volatile carbon and oxygen from the outer disk into the atmosphere of an accreting giant planet, providing a new mechanism to set the C/O and metallicity of giant planet atmospheres separate distinct from planetesimal accretion models. 

\subsection{Heating from the planetary embryo}

We emphasize that DRIVE does not require solids to filter through the gap, nor pass within the envelope of the embedded planet. Previous studies have shown that pressure traps may be ``leaky'' and small particles may filter from the outer disk to the inner disk \citep{dullemond_disk_2018, weber_characterizing_2018, haugbolle_probing_2019, stammler_leaky_2023, price_dynamics_2025, van_clepper_three-dimensional_2025}. In doing so, they may pass by the luminous, embedded planet \citep{petrovic_material_2024}, heating those grains and sublimating ice back to the disk \citep{barnett_thermal_2022, jiang_chemical_2023, wang_atmospheric_2023, cridland_gas_2025}.
Additionally, in 3D, accreted grains tend to pass near the planet following the gas accretion flow, which transports grains near the surface of the disk \citep{szulagyi_meridional_2022, petrovic_material_2024, van_clepper_three-dimensional_2025}. As a result, even when grains are accreted onto the planet, some amount of thermal processing, including sublimation of the ice mantle, is likely to occur during this accretion phase. In the DRIVE mechanism presented here, solids are heated by irradiation from the host star at the atmosphere of the disk rather than from the planet itself, and thus is not dependent on the amount of solid mass that may pass the growing the planet. In this way, the DRIVE mechanism does not depend on the ``leakiness'' of the gap, and is expected to function in either strong or weak pressure bumps.

\section{Conclusion}
In this work, we present a novel method to enrich the atmospheres of cold giant planets with volatiles, increasing their bulk metallicity. This ``Dust Recycling and Icy Volatile Enrichment'' (DRIVE) effect relies on the following disk processes to occur:
\begin{enumerate}
    \item Dust and pebbles are concentrated at the local pressure maxima exterior to the gap carved by a giant planet due to inward drift of icy pebbles from the outer disk.
    \item Small dust is created via pebble fragmentation, which is transported upwards through diffusion and advection via the meridional flows created by the planet and exposed to higher temperatures at the surface of the disk, resulting in the sublimation of volatiles into the gas phase.
    \item The giant planet accretes gas from above one scale height in the disk, which has been enriched in volatiles sublimated from the small dust.
\end{enumerate}

All of these are expected via the current understanding of giant planet formation, and combined will enrich giant planet atmospheres with volatiles, even after the planet has reached the pebble isolation mass and outside the midplane snowline. Similar to how radial drift can enrich volatile abundances interior to midplane snowlines, we expect that this combined fragmentation and lofting should increase volatile abundances above the snow surface as well. While some of the sublimated volatiles may diffuse downward and re-freeze onto grain surfaces, these grains will continue to cycle back to the upper region of the disk, sublimating volatiles again. For the example considered here, this concentration of dust in the pressure bump may lead to enhancements of up to 3$\times$ Solar abundance of CO above the snow surface. This late stage metallicity enhancement in the accretion of giant planets may explain the recently shown relationship between mass and metallicity, indicating a super-solar metallicity in even the most massive giant planets \citep{chachan_revising_2025}.

Under the DRIVE mechanism, we expect wide separation giant planets to become enriched in volatile C and O, increasing measured metallicity in their atmospheres. This effect may explain the observed mismatch between gas-phase composition of disks and atmospheric composition of cold giant planets, which are currently difficult to explain using either pebble or planetesimal accretion models \citep{bergin_co_2024, nasedkin_four---kind_2024, balmer_jwst-tst_2025}. Additionally, DRIVE provides a novel way to separate volatile ice from refractory carriers. Thus, bulk metallicity measurements of exoplanets inferred from single molecular abundances may not reflect this partitioning, and refractory-to-volatile ratios may be key in understanding giant planet formation history \citep{lothringer_new_2021, lothringer_refractory_2025}. Jupiter's enhanced metallicity, for example, is largely based on the enhancement of volatile species, including noble gasses, carbon, nitrogen, and oxygen \citep[see,][and references therein]{atreya_deep_2020}. Understanding the relative composition of refractory elements will be key to understanding Jupiter's bulk metallicity and differentiating possible origin scenarios.

While we have explored the dust growth and fragmentation, disk hydrodynamics, radiative transfer, and particle dynamics separately here, there are important feedback mechanisms that may affect the magnitude of the DRIVE effect. The chemistry of the disk is sensitive to the small dust, which is the primary carrier of volatiles, which may act to further alter the C/O and metallicity of the gas near the growing planet. This dust, however, is intricately linked to the gas dynamics, which is in turn driven by interactions with the embedded planet. This small dust also serves as the primary source of opacity in the disk, feeding back on the thermal structure and UV driven chemistry in the disk. Better understanding the evolution of dust in a disk containing an embedded planet will provide many future avenues of research with implications for the chemistry and dynamics of disks and the planets that form within.

\section*{Acknowledgments}
\added{We thank the anonymous reviewer for their helpful feedback in the preparation of this manuscript.}
This work was completed in part with resources provided by the University of Chicago’s Research Computing Center.
EVC acknowledges support from NASA FINESST grant 80NSSC23K1380. FA is funded by the European Union (ERC, UNVEIL, 101076613). Views and opinions expressed are however those of the author(s) only and do not necessarily reflect those of the European Union or the European Research Council. Neither the European Union nor the granting authority can be held responsible for them.  FJC and EAB acknowledge support from NASA's Emerging Worlds Program, grant 80NSSC20K0333, and NASA's Exoplanets Research Program, 80NSSC20K0259.  This material is also based upon work supported by the National Aeronautics and Space Administration under agreement No. 80NSSC21K0593 for the program “Alien Earths.” The results reported herein benefited from collaborations and/or information exchange within NASA's Nexus for Exoplanet System Science (NExSS) research coordination network sponsored by NASA's Science Mission Directorate.

\software{
\texttt{DustPy} \citep{stammler_dustpy_2022},
\texttt{FARGO3D} \citep{benitez-llambay_fargo3d_2016, masset_fargo_2000},
\texttt{matplotlib} \citep{hunter_matplotlib_2007},
\texttt{numpy} \citep{harris_array_2020}.
\texttt{radmc3d-2.0} \citep{dullemond_radmc-3d_2012}.
}

\bibliography{references}

@ARTICLE{price_dynamics_2025,
       author = {{Price}, Ellen M. and {Van Clepper}, Eric and {Ciesla}, Fred J.},
        title = "{Dynamics of Small, Constant-size Particles in a Protoplanetary Disk with an Embedded Protoplanet}",
      journal = {\apj},
         year = 2025,
        month = jan,
       volume = {979},
       number = {1},
          eid = {37},
        pages = {37},
          doi = {10.3847/1538-4357/ad9f35},
archivePrefix = {arXiv},
       eprint = {2501.17232},
 primaryClass = {astro-ph.EP},
       adsurl = {https://ui.adsabs.harvard.edu/abs/2025ApJ...979...37P}
}

@ARTICLE{bailey_growing_2024,
       author = {{Bailey}, Avery P. and {Zhu}, Zhaohuan},
        title = "{Growing planet envelopes in spite of recycling flows}",
      journal = {\mnras},
         year = 2024,
        month = nov,
       volume = {534},
       number = {3},
        pages = {2953-2967},
          doi = {10.1093/mnras/stae2250},
archivePrefix = {arXiv},
       eprint = {2310.03117},
 primaryClass = {astro-ph.EP},
       adsurl = {https://ui.adsabs.harvard.edu/abs/2024MNRAS.534.2953B}
}

@ARTICLE{chachan_revising_2025,
       author = {{Chachan}, Yayaati and {Fortney}, Jonathan J. and {Ohno}, Kazumasa and {Thorngren}, Daniel and {Murray-Clay}, Ruth},
        title = "{Revising the Giant Planet Mass-Metallicity Relation: Deciphering the Formation Sequence of Giant Planets}",
      journal = {arXiv e-prints},
         year = 2025,
        month = sep,
          eid = {arXiv:2509.20428},
        pages = {arXiv:2509.20428},
          doi = {10.48550/arXiv.2509.20428},
archivePrefix = {arXiv},
       eprint = {2509.20428},
 primaryClass = {astro-ph.EP},
       adsurl = {https://ui.adsabs.harvard.edu/abs/2025arXiv250920428C}
}

@ARTICLE{kurokawa_suppression_2018,
       author = {{Kurokawa}, Hiroyuki and {Tanigawa}, Takayuki},
        title = "{Suppression of atmospheric recycling of planets embedded in a protoplanetary disc by buoyancy barrierc}",
      journal = {\mnras},
         year = 2018,
        month = sep,
       volume = {479},
       number = {1},
        pages = {635-648},
          doi = {10.1093/mnras/sty1498},
archivePrefix = {arXiv},
       eprint = {1806.01695},
 primaryClass = {astro-ph.EP},
       adsurl = {https://ui.adsabs.harvard.edu/abs/2018MNRAS.479..635K}
}

@article{thorngren_MASS_2016,
  title = {{{THE MASS}}--{{METALLICITY RELATION FOR GIANT PLANETS}}},
  author = {Thorngren, Daniel P. and Fortney, Jonathan J. and {Murray-Clay}, Ruth A. and Lopez, Eric D.},
  year = {2016},
  month = nov,
  journal = {ApJ},
  volume = {831},
  number = {1},
  pages = {64},
  issn = {0004-637X, 1538-4357},
  doi = {10.3847/0004-637X/831/1/64},
  urldate = {2025-10-10},
  langid = {english}
}

@ARTICLE{li_carbon_2021,
       author = {{Li}, J. and {Bergin}, E.~A. and {Blake}, G.~A. and {Ciesla}, F.~J. and {Hirschmann}, M.~M.},
        title = "{Earth's carbon deficit caused by early loss through irreversible sublimation}",
      journal = {Science Advances},
         year = 2021,
        month = apr,
       volume = {7},
       number = {14},
        pages = {eabd3632},
          doi = {10.1126/sciadv.abd3632},
archivePrefix = {arXiv},
       eprint = {2104.02702},
 primaryClass = {astro-ph.EP},
       adsurl = {https://ui.adsabs.harvard.edu/abs/2021SciA....7.3632L}
}

@ARTICLE{lee_fire_2010,
       author = {{Lee}, Jeong-Eun and {Bergin}, Edwin A. and {Nomura}, Hideko},
        title = "{The Solar Nebula on Fire: A Solution to the Carbon Deficit in the Inner Solar System}",
      journal = {\apjl},
         year = 2010,
        month = feb,
       volume = {710},
       number = {1},
        pages = {L21-L25},
          doi = {10.1088/2041-8205/710/1/L21},
archivePrefix = {arXiv},
       eprint = {1001.0818},
 primaryClass = {astro-ph.GA},
       adsurl = {https://ui.adsabs.harvard.edu/abs/2010ApJ...710L..21L}
}

@ARTICLE{kress_soot_2010,
       author = {{Kress}, Monika E. and {Tielens}, Alexander G.~G.~M. and {Frenklach}, Michael},
        title = "{The {\textquoteleft}soot line{\textquoteright}: Destruction of presolar polycyclic aromatic hydrocarbons in the terrestrial planet-forming region of disks}",
      journal = {Advances in Space Research},
         year = 2010,
        month = jul,
       volume = {46},
       number = {1},
        pages = {44-49},
          doi = {10.1016/j.asr.2010.02.004},
       adsurl = {https://ui.adsabs.harvard.edu/abs/2010AdSpR..46...44K}
}

@ARTICLE{atreya_comparison_1999,
       author = {{Atreya}, S.~K. and {Wong}, M.~H. and {Owen}, T.~C. and {Mahaffy}, P.~R. and {Niemann}, H.~B. and {de Pater}, I. and {Drossart}, P. and {Encrenaz}, Th},
        title = "{A comparison of the atmospheres of Jupiter and Saturn: deep atmospheric composition, cloud structure, vertical mixing, and origin}",
      journal = {\planss},
         year = 1999,
        month = oct,
       volume = {47},
       number = {10-11},
        pages = {1243-1262},
          doi = {10.1016/S0032-0633(99)00047-1},
       adsurl = {https://ui.adsabs.harvard.edu/abs/1999P&SS...47.1243A}
}

@ARTICLE{villenave_turbulence_2025,
       author = {{Villenave}, Marion and {Rosotti}, Giovanni P. and {Lambrechts}, Michiel and {Ziampras}, Alexandros and {Pinte}, Christophe and {M{\'e}nard}, Fran{\c{c}}ois and {Stapelfeldt}, Karl R. and {Duch{\^e}ne}, Gaspard and {Baylock}, Emily and {Doi}, Kiyoaki},
        title = "{Turbulence in protoplanetary disks: A systematic analysis of dust settling in 33 disks}",
      journal = {\aap},
         year = 2025,
        month = may,
       volume = {697},
          eid = {A64},
        pages = {A64},
          doi = {10.1051/0004-6361/202553822},
archivePrefix = {arXiv},
       eprint = {2503.05872},
 primaryClass = {astro-ph.SR},
       adsurl = {https://ui.adsabs.harvard.edu/abs/2025A&A...697A..64V}
}

@ARTICLE{mathis_size_1977,
       author = {{Mathis}, J.~S. and {Rumpl}, W. and {Nordsieck}, K.~H.},
        title = "{The size distribution of interstellar grains.}",
      journal = {\apj},
         year = 1977,
        month = oct,
       volume = {217},
        pages = {425-433},
          doi = {10.1086/155591},
       adsurl = {https://ui.adsabs.harvard.edu/abs/1977ApJ...217..425M}
}

@ARTICLE{calvet_flat_1994,
       author = {{Calvet}, Nuria and {Hartmann}, Lee and {Kenyon}, S.~J. and {Whitney}, B.~A.},
        title = "{Flat Spectrum T Tauri Stars: The Case for Infall}",
      journal = {\apj},
         year = 1994,
        month = oct,
       volume = {434},
        pages = {330},
          doi = {10.1086/174731},
       adsurl = {https://ui.adsabs.harvard.edu/abs/1994ApJ...434..330C}
}

@ARTICLE{danti_composition_2023,
       author = {{Danti}, C. and {Bitsch}, B. and {Mah}, J.},
        title = "{Composition of giant planets: The roles of pebbles and planetesimals}",
      journal = {\aap},
         year = 2023,
        month = nov,
       volume = {679},
          eid = {L7},
        pages = {L7},
          doi = {10.1051/0004-6361/202347501},
archivePrefix = {arXiv},
       eprint = {2310.02886},
 primaryClass = {astro-ph.EP},
       adsurl = {https://ui.adsabs.harvard.edu/abs/2023A&A...679L...7D}
}

@ARTICLE{pollack_formation_1996,
       author = {{Pollack}, James B. and {Hubickyj}, Olenka and {Bodenheimer}, Peter and {Lissauer}, Jack J. and {Podolak}, Morris and {Greenzweig}, Yuval},
        title = "{Formation of the Giant Planets by Concurrent Accretion of Solids and Gas}",
      journal = {\icarus},
         year = 1996,
        month = nov,
       volume = {124},
       number = {1},
        pages = {62-85},
          doi = {10.1006/icar.1996.0190},
       adsurl = {https://ui.adsabs.harvard.edu/abs/1996Icar..124...62P}
}

@ARTICLE{kley_three-dimensional_2001,
       author = {{Kley}, Wilhelm and {D'Angelo}, Gennaro and {Henning}, Thomas},
        title = "{Three-dimensional Simulations of a Planet Embedded in a Protoplanetary Disk}",
      journal = {\apj},
         year = 2001,
        month = jan,
       volume = {547},
       number = {1},
        pages = {457-464},
          doi = {10.1086/318345},
       adsurl = {https://ui.adsabs.harvard.edu/abs/2001ApJ...547..457K}
}

@article{ciesla_evolution_2006,
	title = {The evolution of the water distribution in a viscous protoplanetary disk},
	volume = {181},
	copyright = {https://www.elsevier.com/tdm/userlicense/1.0/},
	issn = {00191035},
	url = {https://linkinghub.elsevier.com/retrieve/pii/S0019103505004641},
	doi = {10.1016/j.icarus.2005.11.009},
	language = {en},
	number = {1},
	urldate = {2025-06-09},
	journal = {Icarus},
	author = {Ciesla, F and Cuzzi, J},
	month = mar,
	year = {2006},
	pages = {178--204},
}

@article{cuzzi_material_2004,
	title = {Material {Enhancement} in {Protoplanetary} {Nebulae} by {Particle} {Drift} through {Evaporation} {Fronts}},
	volume = {614},
	issn = {0004-637X, 1538-4357},
	url = {https://iopscience.iop.org/article/10.1086/423611},
	doi = {10.1086/423611},
	language = {en},
	number = {1},
	urldate = {2025-06-09},
	journal = {The Astrophysical Journal},
	author = {Cuzzi, Jeffrey N. and Zahnle, Kevin J.},
	month = oct,
	year = {2004},
	pages = {490--496},
}

@article{eriksson_low_2022,
	title = {A low accretion efficiency of planetesimals formed at planetary gap edges},
	volume = {661},
	copyright = {https://www.edpsciences.org/en/authors/copyright-and-licensing},
	issn = {0004-6361, 1432-0746},
	url = {https://www.aanda.org/10.1051/0004-6361/202142391},
	doi = {10.1051/0004-6361/202142391},
	language = {en},
	urldate = {2025-06-06},
	journal = {Astronomy \& Astrophysics},
	author = {Eriksson, Linn E. J. and Ronnet, Thomas and Johansen, Anders and Helled, Ravit and Valletta, Claudio and Petit, Antoine C.},
	month = may,
	year = {2022},
	pages = {A73},
}

@article{lothringer_new_2021,
	title = {A {New} {Window} into {Planet} {Formation} and {Migration}: {Refractory}-to-{Volatile} {Elemental} {Ratios} in {Ultra}-hot {Jupiters}},
	language = {en},
	journal = {The Astrophysical Journal},
	author = {Lothringer, Joshua D},
	year = {2021},
}

@article{drazkowska_planetesimal_2017,
	title = {Planetesimal formation starts at the snow line},
	volume = {608},
	issn = {0004-6361, 1432-0746},
	url = {http://www.aanda.org/10.1051/0004-6361/201731491},
	doi = {10.1051/0004-6361/201731491},
	language = {en},
	urldate = {2025-06-04},
	journal = {Astronomy \& Astrophysics},
	author = {Drążkowska, J. and Alibert, Y.},
	month = dec,
	year = {2017},
	pages = {A92},
}

@article{ciesla_organic_2012,
	title = {Organic {Synthesis} via {Irradiation} and {Warming} of {Ice} {Grains} in the {Solar} {Nebula}},
	volume = {336},
	issn = {0036-8075, 1095-9203},
	url = {https://www.science.org/doi/10.1126/science.1217291},
	doi = {10.1126/science.1217291},
	language = {en},
	number = {6080},
	urldate = {2025-06-03},
	journal = {Science},
	author = {Ciesla, Fred J. and Sandford, Scott A.},
	month = apr,
	year = {2012},
	pages = {452--454},
}

@article{fayolle_n2_2016,
	title = {N$_{\textrm{2}}$ {AND} {CO} {DESORPTION} {ENERGIES} {FROM} {WATER} {ICE}},
	volume = {816},
	issn = {2041-8205, 2041-8213},
	url = {https://iopscience.iop.org/article/10.3847/2041-8205/816/2/L28},
	doi = {10.3847/2041-8205/816/2/L28},
	language = {en},
	number = {2},
	urldate = {2025-05-29},
	journal = {The Astrophysical Journal Letters},
	author = {Fayolle, Edith C. and Balfe, Jodi and Loomis, Ryan and Bergner, Jennifer and Graninger, Dawn and Rajappan, Mahesh and Öberg, Karin I.},
	month = jan,
	year = {2016},
	pages = {L28},
}

@article{li_water_2020,
	title = {The water abundance in {Jupiter}’s equatorial zone},
	volume = {4},
	issn = {2397-3366},
	url = {https://www.nature.com/articles/s41550-020-1009-3},
	doi = {10.1038/s41550-020-1009-3},
	language = {en},
	number = {6},
	urldate = {2025-05-29},
	journal = {Nature Astronomy},
	author = {Li, Cheng and Ingersoll, Andrew and Bolton, Scott and Levin, Steven and Janssen, Michael and Atreya, Sushil and Lunine, Jonathan and Steffes, Paul and Brown, Shannon and Guillot, Tristan and Allison, Michael and Arballo, John and Bellotti, Amadeo and Adumitroaie, Virgil and Gulkis, Samuel and Hodges, Amoree and Li, Liming and Misra, Sidharth and Orton, Glenn and Oyafuso, Fabiano and Santos-Costa, Daniel and Waite, Hunter and Zhang, Zhimeng},
	month = feb,
	year = {2020},
	pages = {609--616},
}

@article{atreya_deep_2020,
	title = {Deep {Atmosphere} {Composition}, {Structure}, {Origin}, and {Exploration}, with {Particular} {Focus} on {Critical} in situ {Science} at the {Icy} {Giants}},
	volume = {216},
	issn = {0038-6308, 1572-9672},
	url = {http://link.springer.com/10.1007/s11214-020-0640-8},
	doi = {10.1007/s11214-020-0640-8},
	language = {en},
	number = {1},
	urldate = {2025-05-29},
	journal = {Space Science Reviews},
	author = {Atreya, Sushil K. and Hofstadter, Mark H. and In, Joong Hyun and Mousis, Olivier and Reh, Kim and Wong, Michael H.},
	month = feb,
	year = {2020},
	pages = {18},
}

@article{monga_external_2014,
	title = {{EXTERNAL} {PHOTOEVAPORATION} {OF} {THE} {SOLAR} {NEBULA}: {JUPITER}'s {NOBLE} {GAS} {ENRICHMENTS}},
	volume = {798},
	copyright = {http://iopscience.iop.org/info/page/text-and-data-mining},
	issn = {1538-4357},
	shorttitle = {{EXTERNAL} {PHOTOEVAPORATION} {OF} {THE} {SOLAR} {NEBULA}},
	url = {https://iopscience.iop.org/article/10.1088/0004-637X/798/1/9},
	doi = {10.1088/0004-637X/798/1/9},
	language = {en},
	number = {1},
	urldate = {2025-05-29},
	journal = {The Astrophysical Journal},
	author = {Monga, Nikhil and Desch, Steven},
	month = dec,
	year = {2014},
	pages = {9},
}

@article{mousis_jupiters_2019,
	title = {Jupiter’s {Formation} in the {Vicinity} of the {Amorphous} {Ice} {Snowline}},
	volume = {875},
	issn = {0004-637X, 1538-4357},
	url = {https://iopscience.iop.org/article/10.3847/1538-4357/ab0a72},
	doi = {10.3847/1538-4357/ab0a72},
	language = {en},
	number = {1},
	urldate = {2025-05-29},
	journal = {The Astrophysical Journal},
	author = {Mousis, Olivier and Ronnet, Thomas and Lunine, Jonathan I.},
	month = apr,
	year = {2019},
	pages = {9},
}

@article{meijerink_radiative_2009,
	title = {{RADIATIVE} {TRANSFER} {MODELS} {OF} {MID}-{INFRARED} {H}$_{\textrm{2}}$ {O} {LINES} {IN} {THE} {PLANET}-{FORMING} {REGION} {OF} {CIRCUMSTELLAR} {DISKS}},
	volume = {704},
	issn = {0004-637X, 1538-4357},
	url = {https://iopscience.iop.org/article/10.1088/0004-637X/704/2/1471},
	doi = {10.1088/0004-637X/704/2/1471},
	language = {en},
	number = {2},
	urldate = {2025-05-12},
	journal = {The Astrophysical Journal},
	author = {Meijerink, R. and Pontoppidan, K. M. and Blake, G. A. and Poelman, D. R. and Dullemond, C. P.},
	month = oct,
	year = {2009},
	pages = {1471--1481},
}

@article{booth_chemical_2017,
	title = {Chemical enrichment of giant planets and discs due to pebble drift},
	volume = {469},
	issn = {0035-8711, 1365-2966},
	url = {https://academic.oup.com/mnras/article-lookup/doi/10.1093/mnras/stx1103},
	doi = {10.1093/mnras/stx1103},
	language = {en},
	number = {4},
	urldate = {2025-05-09},
	journal = {Monthly Notices of the Royal Astronomical Society},
	author = {Booth, Richard A. and Clarke, Cathie J. and Madhusudhan, Nikku and Ilee, John D.},
	month = aug,
	year = {2017},
	pages = {3994--4011},
}

@article{zhang_excess_2020,
	title = {Excess {C}/{H} in {Protoplanetary} {Disk} {Gas} from {Icy} {Pebble} {Drift} {Across} the {CO} {Snowline}},
	volume = {891},
	issn = {2041-8205, 2041-8213},
	url = {https://iopscience.iop.org/article/10.3847/2041-8213/ab77ca},
	doi = {10.3847/2041-8213/ab77ca},
	language = {en},
	number = {1},
	urldate = {2025-05-09},
	journal = {The Astrophysical Journal Letters},
	author = {Zhang, Ke and Bosman, Arthur D. and Bergin, Edwin A.},
	month = mar,
	year = {2020},
	pages = {L16},
}

@article{oberg_excess_2016,
	title = {{EXCESS} {C}/{O} {AND} {C}/{H} {IN} {OUTER} {PROTOPLANETARY} {DISK} {GAS}},
	volume = {831},
	issn = {2041-8205, 2041-8213},
	url = {https://iopscience.iop.org/article/10.3847/2041-8205/831/2/L19},
	doi = {10.3847/2041-8205/831/2/L19},
	language = {en},
	number = {2},
	urldate = {2025-05-09},
	journal = {The Astrophysical Journal Letters},
	author = {Öberg, Karin I. and Bergin, Edwin A.},
	month = nov,
	year = {2016},
	pages = {L19},
}

@misc{dullemond_radmc-3d_2012,
	title = {{RADMC}-{3D}},
	isbn = {ascl:1202.015},
	url = {https://ascl.net/1202.015},
	author = {Dullemond, C P and Juhasz, A. and Pohl, A and Seresheti, F and Shetty, R and Peters, T and Commercon, B and Flock, M},
	year = {2012},
}

@article{bi_puffed-up_2021,
	title = {Puffed-up {Edges} of {Planet}-opened {Gaps} in {Protoplanetary} {Disks}. {I}. {Hydrodynamic} {Simulations}},
	volume = {912},
	issn = {0004-637X, 1538-4357},
	url = {https://iopscience.iop.org/article/10.3847/1538-4357/abef6b},
	doi = {10.3847/1538-4357/abef6b},
	language = {en},
	number = {2},
	urldate = {2025-03-10},
	journal = {The Astrophysical Journal},
	author = {Bi, Jiaqing and Lin, Min-Kai and Dong, Ruobing},
	month = may,
	year = {2021},
	pages = {107},
}

@ARTICLE{wang_atmospheric_2023,
       author = {{Wang}, Yu and {Ormel}, Chris W. and {Huang}, Pinghui and {Kuiper}, Rolf},
        title = "{Atmospheric recycling of volatiles by pebble-accreting planets}",
      journal = {\mnras},
         year = 2023,
        month = aug,
       volume = {523},
       number = {4},
        pages = {6186-6207},
          doi = {10.1093/mnras/stad1753},
archivePrefix = {arXiv},
       eprint = {2306.06169},
 primaryClass = {astro-ph.EP},
       adsurl = {https://ui.adsabs.harvard.edu/abs/2023MNRAS.523.6186W}
}

@article{shibata_origin_2022,
	title = {The origin of the high metallicity of close-in giant exoplanets: {II}. {The} nature of the sweet spot for accretion},
	volume = {659},
	copyright = {https://www.edpsciences.org/en/authors/copyright-and-licensing},
	issn = {0004-6361, 1432-0746},
	shorttitle = {The origin of the high metallicity of close-in giant exoplanets},
	url = {https://www.aanda.org/10.1051/0004-6361/202142180},
	doi = {10.1051/0004-6361/202142180},
	language = {en},
	urldate = {2025-03-27},
	journal = {Astronomy \& Astrophysics},
	author = {Shibata, S. and Helled, R. and Ikoma, M.},
	month = mar,
	year = {2022},
	pages = {A28},
}

@article{shibata_origin_2020,
	title = {The origin of the high metallicity of close-in giant exoplanets: {Combined} effects of resonant and aerodynamic shepherding},
	volume = {633},
	copyright = {https://www.edpsciences.org/en/authors/copyright-and-licensing},
	issn = {0004-6361, 1432-0746},
	shorttitle = {The origin of the high metallicity of close-in giant exoplanets},
	url = {https://www.aanda.org/10.1051/0004-6361/201936700},
	doi = {10.1051/0004-6361/201936700},
	language = {en},
	urldate = {2025-03-27},
	journal = {Astronomy \& Astrophysics},
	author = {Shibata, Sho and Helled, Ravit and Ikoma, Masahiro},
	month = jan,
	year = {2020},
	pages = {A33},
}

@article{shibata_enrichment_2022,
	title = {Enrichment of {Jupiter}’s {Atmosphere} by {Late} {Planetesimal} {Bombardment}},
	volume = {926},
	issn = {2041-8205, 2041-8213},
	url = {https://iopscience.iop.org/article/10.3847/2041-8213/ac54b1},
	doi = {10.3847/2041-8213/ac54b1},
	language = {en},
	number = {2},
	urldate = {2025-03-27},
	journal = {The Astrophysical Journal Letters},
	author = {Shibata, Sho and Helled, Ravit},
	month = feb,
	year = {2022},
	pages = {L37},
}

@article{hsu_pds_2024,
	title = {{PDS} 70b {Shows} {Stellar}-like {Carbon}-to-oxygen {Ratio}},
	volume = {977},
	issn = {2041-8205, 2041-8213},
	url = {https://iopscience.iop.org/article/10.3847/2041-8213/ad95e8},
	doi = {10.3847/2041-8213/ad95e8},
	language = {en},
	number = {2},
	urldate = {2025-03-20},
	journal = {The Astrophysical Journal Letters},
	author = {Hsu, Chih-Chun and Wang, Jason J. and Blake, Geoffrey A. and Xuan, Jerry W. and Zhang, Yapeng and Ruffio, Jean-Baptiste and Horstman, Katelyn and Cronin, Julianne and Sappey, Ben and Xin, Yinzi and Finnerty, Luke and Echeverri, Daniel and Mawet, Dimitri and Jovanovic, Nemanja and Do Ó, Clarissa R. and Baker, Ashley and Bartos, Randall and Calvin, Benjamin and Cetre, Sylvain and Delorme, Jacques-Robert and Doppmann, Gregory W. and Fitzgerald, Michael P. and Liberman, Joshua and López, Ronald A. and Morris, Evan and Pezzato-Rovner, Jacklyn and Schofield, Tobias and Skemer, Andrew and Wallace, J. Kent and Wang 王, Ji 吉},
	month = dec,
	year = {2024},
	pages = {L47},
}

@article{andama_role_2022,
	title = {The role of density perturbation on planet formation by pebble accretion},
	volume = {512},
	copyright = {https://academic.oup.com/journals/pages/open\_access/funder\_policies/chorus/standard\_publication\_model},
	issn = {0035-8711, 1365-2966},
	url = {https://academic.oup.com/mnras/article/512/4/5278/6552143},
	doi = {10.1093/mnras/stac772},
	language = {en},
	number = {4},
	urldate = {2025-03-20},
	journal = {Monthly Notices of the Royal Astronomical Society},
	author = {Andama, G and Ndugu, N and Anguma, S K and Jurua, E},
	month = apr,
	year = {2022},
	pages = {5278--5297},
}

@article{lau_rapid_2022,
	title = {Rapid formation of massive planetary cores in a pressure bump},
	volume = {668},
	copyright = {https://creativecommons.org/licenses/by/4.0},
	issn = {0004-6361, 1432-0746},
	url = {https://www.aanda.org/10.1051/0004-6361/202244864},
	doi = {10.1051/0004-6361/202244864},
	language = {en},
	urldate = {2025-03-20},
	journal = {Astronomy \& Astrophysics},
	author = {Lau, Tommy Chi Ho and Drążkowska, Joanna and Stammler, Sebastian M. and Birnstiel, Tilman and Dullemond, Cornelis P.},
	month = dec,
	year = {2022},
	pages = {A170},
}

@article{bergin_co_2024,
	title = {C/{O} {Ratios} and the {Formation} of {Wide}-separation {Exoplanets}},
	volume = {969},
	issn = {2041-8205, 2041-8213},
	url = {https://iopscience.iop.org/article/10.3847/2041-8213/ad5839},
	doi = {10.3847/2041-8213/ad5839},
	language = {en},
	number = {1},
	urldate = {2025-03-20},
	journal = {The Astrophysical Journal Letters},
	author = {Bergin, Edwin A. and Booth, Richard A. and Colmenares, Maria Jose and Ilee, John D.},
	month = jul,
	year = {2024},
	pages = {L21},
}

@misc{lothringer_refractory_2025,
	title = {Refractory and {Volatile} {Species} in the {UV}-to-{IR} {Transmission} {Spectrum} of {Ultra}-hot {Jupiter} {WASP}-178b with {HST} and {JWST}},
	url = {http://arxiv.org/abs/2503.15472},
	doi = {10.3847/1538-3881/adc117},
	language = {en},
	urldate = {2025-03-20},
	author = {Lothringer, Joshua D. and Bennett, Katherine A. and Sing, David K. and Kehoe-Seamons, Brian and Rustamkulov, Zafar and Reggiani, Henrique and Schlaufman, Kevin C. and McCreery, Patrick and Norris, Seti and Hauschildt, Peter and Cacho-Negrete, Ceiligh and Gressier, Amélie and Espinoza, Néstor and Gapp, Cyril and Evans-Soma, Thomas M. and Stevenson, Kevin B. and Wakeford, Hannah R. and Gibson, Neale and Wilson, Jamie and Nikolov, Nikolay},
	month = mar,
	year = {2025},
	note = {arXiv:2503.15472 [astro-ph]},
}

@article{nasedkin_four---kind_2024,
	title = {Four-of-a-kind? {Comprehensive} atmospheric characterisation of the {HR} 8799 planets with {VLTI}/{GRAVITY}},
	volume = {687},
	copyright = {https://creativecommons.org/licenses/by/4.0},
	issn = {0004-6361, 1432-0746},
	shorttitle = {Four-of-a-kind?},
	url = {https://www.aanda.org/10.1051/0004-6361/202449328},
	doi = {10.1051/0004-6361/202449328},
	language = {en},
	urldate = {2025-03-20},
	journal = {Astronomy \& Astrophysics},
	author = {Nasedkin, E. and Mollière, P. and Lacour, S. and Nowak, M. and Kreidberg, L. and Stolker, T. and Wang, J. J. and Balmer, W. O. and Kammerer, J. and Shangguan, J. and Abuter, R. and Amorim, A. and Asensio-Torres, R. and Benisty, M. and Berger, J.-P. and Beust, H. and Blunt, S. and Boccaletti, A. and Bonnefoy, M. and Bonnet, H. and Bordoni, M. S. and Bourdarot, G. and Brandner, W. and Cantalloube, F. and Caselli, P. and Charnay, B. and Chauvin, G. and Chavez, A. and Choquet, E. and Christiaens, V. and Clénet, Y. and Coudé Du Foresto, V. and Cridland, A. and Davies, R. and Dembet, R. and Dexter, J. and Drescher, A. and Duvert, G. and Eckart, A. and Eisenhauer, F. and Förster Schreiber, N. M. and Garcia, P. and Garcia Lopez, R. and Gendron, E. and Genzel, R. and Gillessen, S. and Girard, J. H. and Grant, S. and Haubois, X. and Heißel, G. and Henning, Th. and Hinkley, S. and Hippler, S. and Houllé, M. and Hubert, Z. and Jocou, L. and Keppler, M. and Kervella, P. and Kurtovic, N. T. and Lagrange, A.-M. and Lapeyrère, V. and Le Bouquin, J.-B. and Lutz, D. and Maire, A.-L. and Mang, F. and Marleau, G.-D. and Mérand, A. and Monnier, J. D. and Mordasini, C. and Ott, T. and Otten, G. P. P. L. and Paladini, C. and Paumard, T. and Perraut, K. and Perrin, G. and Pfuhl, O. and Pourré, N. and Pueyo, L. and Ribeiro, D. C. and Rickman, E. and Ruffio, J. B. and Rustamkulov, Z. and Shimizu, T. and Sing, D. and Stadler, J. and Straub, O. and Straubmeier, C. and Sturm, E. and Tacconi, L. J. and Van Dishoeck, E. F. and Vigan, A. and Vincent, F. and Von Fellenberg, S. D. and Widmann, F. and Winterhalder, T. O. and Woillez, J. and Yazici, Ş. and {the GRAVITY Collaboration}},
	month = jul,
	year = {2024},
	pages = {A298},
}

@article{balmer_jwst-tst_2025,
	title = {{JWST}-{TST} {High} {Contrast}: {Living} on the {Wedge}, or, {NIRCam} {Bar} {Coronagraphy} {Reveals} {CO}$_{\textrm{2}}$ in the {HR} 8799 and 51 {Eri} {Exoplanets}’ {Atmospheres}},
	volume = {169},
	issn = {0004-6256, 1538-3881},
	shorttitle = {{JWST}-{TST} {High} {Contrast}},
	url = {https://iopscience.iop.org/article/10.3847/1538-3881/adb1c6},
	doi = {10.3847/1538-3881/adb1c6},
	language = {en},
	number = {4},
	urldate = {2025-03-19},
	journal = {The Astronomical Journal},
	author = {Balmer, William O. and Kammerer, Jens and Pueyo, Laurent and Perrin, Marshall D. and Girard, Julien H. and Leisenring, Jarron M. and Lawson, Kellen and Dennen, Henry and Van Der Marel, Roeland P. and Beichman, Charles A. and Bryden, Geoffrey and Llop-Sayson, Jorge and Valenti, Jeff A. and Lothringer, Joshua D. and Lewis, Nikole K. and Mâlin, Mathilde and Rebollido, Isabel and Rickman, Emily and Hoch, Kielan K. W. and Soummer, Rémi and Clampin, Mark and Mountain, C. Matt},
	month = apr,
	year = {2025},
	pages = {209},
}

@article{lambrechts_forming_2014,
	title = {Forming the cores of giant planets from the radial pebble flux in protoplanetary discs},
	volume = {572},
	issn = {0004-6361, 1432-0746},
	url = {http://www.aanda.org/10.1051/0004-6361/201424343},
	doi = {10.1051/0004-6361/201424343},
	language = {en},
	urldate = {2025-03-17},
	journal = {Astronomy \& Astrophysics},
	author = {Lambrechts, M. and Johansen, A.},
	month = dec,
	year = {2014},
	pages = {A107},
}

@article{jiang_chemical_2023,
	title = {Chemical footprints of giant planet formation: {Role} of planet accretion in shaping the {C}/{O} ratio of protoplanetary disks},
	volume = {678},
	copyright = {https://creativecommons.org/licenses/by/4.0},
	issn = {0004-6361, 1432-0746},
	shorttitle = {Chemical footprints of giant planet formation},
	url = {https://www.aanda.org/10.1051/0004-6361/202346637},
	doi = {10.1051/0004-6361/202346637},
	language = {en},
	urldate = {2025-03-06},
	journal = {Astronomy \& Astrophysics},
	author = {Jiang, Haochang and Wang, Yu and Ormel, Chris W. and Krijt, Sebastiaan and Dong, Ruobing},
	month = oct,
	year = {2023},
	pages = {A33},
}

@article{krijt_dust_2016,
	title = {{DUST} {DIFFUSION} {AND} {SETTLING} {IN} {THE} {PRESENCE} {OF} {COLLISIONS}: {TRAPPING} ({SUB}){MICRON} {GRAINS} {IN} {THE} {MIDPLANE}},
	volume = {822},
	issn = {0004-637X, 1538-4357},
	shorttitle = {{DUST} {DIFFUSION} {AND} {SETTLING} {IN} {THE} {PRESENCE} {OF} {COLLISIONS}},
	url = {https://iopscience.iop.org/article/10.3847/0004-637X/822/2/111},
	doi = {10.3847/0004-637X/822/2/111},
	language = {en},
	number = {2},
	urldate = {2025-03-03},
	journal = {The Astrophysical Journal},
	author = {Krijt, Sebastiaan and Ciesla, Fred J.},
	month = may,
	year = {2016},
	pages = {111},
}

@article{van_clepper_three-dimensional_2025,
	title = {Three-dimensional {Transport} of {Solids} in a {Protoplanetary} {Disk} {Containing} a {Growing} {Giant} {Planet}},
	volume = {980},
	doi = {10.3847/1538-4357/ada8a4},
	language = {en},
	journal = {The Astrophysical Journal},
	author = {Van Clepper, Eric and Price, Ellen and Ciesla, Fred J.},
	year = {2025},
	pages = {201},
}

@article{lega_gas_2024,
	title = {Gas dynamics around a {Jupiter}-mass planet: {I}. {Influence} of protoplanetary disk properties},
	volume = {690},
	copyright = {https://creativecommons.org/licenses/by/4.0},
	issn = {0004-6361, 1432-0746},
	shorttitle = {Gas dynamics around a {Jupiter}-mass planet},
	url = {https://www.aanda.org/10.1051/0004-6361/202450899},
	doi = {10.1051/0004-6361/202450899},
	language = {en},
	urldate = {2025-01-20},
	journal = {Astronomy \& Astrophysics},
	author = {Lega, E. and Benisty, M. and Cridland, A. and Morbidelli, A. and Schulik, M. and Lambrechts, M.},
	month = oct,
	year = {2024},
	pages = {A183},
}

@article{cridland_gas_2025,
	title = {Gas dynamics around a {Jupiter}-mass planet: {II}. {Chemical} evolution of circumplanetary material},
	volume = {693},
	copyright = {https://creativecommons.org/licenses/by/4.0},
	issn = {0004-6361, 1432-0746},
	shorttitle = {Gas dynamics around a {Jupiter}-mass planet},
	url = {https://www.aanda.org/10.1051/0004-6361/202451140},
	doi = {10.1051/0004-6361/202451140},
	language = {en},
	urldate = {2025-01-20},
	journal = {Astronomy \& Astrophysics},
	author = {Cridland, Alex J. and Lega, Elena and Benisty, Myriam},
	month = jan,
	year = {2025},
	pages = {A86},
}

@misc{flores-rivera_uv-processing_2024,
	title = {{UV}-processing of icy pebbles in the outer parts of {VSI}-turbulent disks},
	url = {http://arxiv.org/abs/2412.01698},
	doi = {10.48550/arXiv.2412.01698},
	language = {en},
	urldate = {2024-12-09},
	publisher = {arXiv},
	author = {Flores-Rivera, Lizxandra and Lambrechts, Michiel and Gavino, Sacha and Lorek, Sebastian and Flock, Mario and Johansen, Anders and Mignone, Andrea},
	month = dec,
	year = {2024},
	note = {arXiv:2412.01698 [astro-ph]},
}

@article{eriksson_particle_2024,
	title = {Particle fragmentation inside planet-induced spiral waves},
	copyright = {https://creativecommons.org/licenses/by/4.0/},
	issn = {1745-3925, 1745-3933},
	url = {https://academic.oup.com/mnrasl/advance-article/doi/10.1093/mnrasl/slae110/7905883},
	doi = {10.1093/mnrasl/slae110},
	language = {en},
	urldate = {2024-12-06},
	journal = {Monthly Notices of the Royal Astronomical Society: Letters},
	author = {Eriksson, Linn E J and Yang, Chao-Chin and Armitage, Philip J},
	month = nov,
	year = {2024},
	pages = {slae110},
}

@article{alarcon_thermal_2024,
	title = {Thermal {Structure} and {Millimeter} {Emission} from a {Protoplanetary} {Disk} with {Embedded} {Protoplanets} from {Radiative} {Transfer} {Modeling}},
	volume = {967},
	issn = {0004-637X, 1538-4357},
	url = {https://iopscience.iop.org/article/10.3847/1538-4357/ad3d57},
	doi = {10.3847/1538-4357/ad3d57},
	language = {en},
	number = {2},
	urldate = {2024-10-11},
	journal = {The Astrophysical Journal},
	author = {Alarcón, Felipe and Bergin, Edwin A.},
	month = jun,
	year = {2024},
	pages = {144},
}

@article{stammler_dustpy_2022,
	title = {{DustPy}: {A} {Python} {Package} for {Dust} {Evolution} in {Protoplanetary} {Disks}},
	volume = {935},
	issn = {0004-637X, 1538-4357},
	shorttitle = {{DustPy}},
	url = {https://iopscience.iop.org/article/10.3847/1538-4357/ac7d58},
	doi = {10.3847/1538-4357/ac7d58},
	language = {en},
	number = {1},
	urldate = {2024-10-07},
	journal = {The Astrophysical Journal},
	author = {Stammler, Sebastian M. and Birnstiel, Tilman},
	month = aug,
	year = {2022},
	pages = {35},
}

@article{maeda_delivery_2024,
	title = {Delivery of {Dust} {Particles} from {Protoplanetary} {Disks} onto {Circumplanetary} {Disks} of {Giant} {Planets}},
	volume = {968},
	issn = {0004-637X, 1538-4357},
	url = {https://iopscience.iop.org/article/10.3847/1538-4357/ad4035},
	doi = {10.3847/1538-4357/ad4035},
	language = {en},
	number = {2},
	urldate = {2024-09-26},
	journal = {The Astrophysical Journal},
	author = {Maeda, Natsuho and Ohtsuki, Keiji and Suetsugu, Ryo and Shibaike, Yuhito and Tanigawa, Takayuki and Machida, Masahiro N.},
	month = jun,
	year = {2024},
	pages = {62},
}

@misc{petrovic_material_2024,
	title = {Material {Transport} in {Protoplanetary} {Discs} with {Massive} {Embedded} {Planets}},
	url = {http://arxiv.org/abs/2409.16245},
	language = {en},
	urldate = {2024-09-26},
	publisher = {arXiv},
	author = {Petrovic, Hannah J. and Booth, Richard A. and Clarke, Cathie J.},
	month = sep,
	year = {2024},
	note = {arXiv:2409.16245 [astro-ph]},
}

@article{bosman_jupiter_2019,
	title = {Jupiter formed as a pebble pile around the {N} $_{\textrm{2}}$ ice line},
	volume = {632},
	copyright = {https://www.edpsciences.org/en/authors/copyright-and-licensing},
	issn = {0004-6361, 1432-0746},
	url = {https://www.aanda.org/10.1051/0004-6361/201936827},
	doi = {10.1051/0004-6361/201936827},
	language = {en},
	urldate = {2024-08-29},
	journal = {Astronomy \& Astrophysics},
	author = {Bosman, A. D. and Cridland, A. J. and Miguel, Y.},
	month = dec,
	year = {2019},
	pages = {L11},
}

@article{dullemond_disk_2018,
	title = {The {Disk} {Substructures} at {High} {Angular} {Resolution} {Project} ({DSHARP}). {VI}. {Dust} {Trapping} in {Thin}-ringed {Protoplanetary} {Disks}},
	volume = {869},
	issn = {2041-8205, 2041-8213},
	url = {https://iopscience.iop.org/article/10.3847/2041-8213/aaf742},
	doi = {10.3847/2041-8213/aaf742},
	language = {en},
	number = {2},
	urldate = {2024-08-13},
	journal = {The Astrophysical Journal Letters},
	author = {Dullemond, Cornelis P. and Birnstiel, Tilman and Huang, Jane and Kurtovic, Nicolás T. and Andrews, Sean M. and Guzmán, Viviana V. and Pérez, Laura M. and Isella, Andrea and Zhu, Zhaohuan and Benisty, Myriam and Wilner, David J. and Bai, Xue-Ning and Carpenter, John M. and Zhang, Shangjia and Ricci, Luca},
	month = dec,
	year = {2018},
	pages = {L46},
}

@article{ligterink_rapid_2024,
	title = {The rapid formation of macromolecules in irradiated ice of protoplanetary disk dust traps},
	issn = {2397-3366},
	url = {https://www.nature.com/articles/s41550-024-02334-4},
	doi = {10.1038/s41550-024-02334-4},
	language = {en},
	urldate = {2024-08-12},
	journal = {Nature Astronomy},
	author = {Ligterink, Niels F. W. and Pinilla, Paola and Van Der Marel, Nienke and Van Scheltinga, Jeroen Terwisscha and Booth, Alice S. and Alexander, Conel M. O’D. and Riebe, My E. I.},
	month = jul,
	year = {2024},
}

@article{oberg_jupiters_2019,
	title = {Jupiter's {Composition} {Suggests} its {Core} {Assembled} {Exterior} to the {N} $_{\textrm{2}}$ {Snowline}},
	volume = {158},
	issn = {0004-6256, 1538-3881},
	url = {https://iopscience.iop.org/article/10.3847/1538-3881/ab46a8},
	doi = {10.3847/1538-3881/ab46a8},
	language = {en},
	number = {5},
	urldate = {2024-08-06},
	journal = {The Astronomical Journal},
	author = {Öberg, Karin I and Wordsworth, Robin},
	month = nov,
	year = {2019},
	pages = {194},
}

@article{doi_estimate_2021,
	title = {Estimate on {Dust} {Scale} {Height} from the {ALMA} {Dust} {Continuum} {Image} of the {HD} 163296 {Protoplanetary} {Disk}},
	language = {en},
	journal = {The Astrophysical Journal},
	author = {Doi, Kiyoaki},
	year = {2021},
}

@article{bitsch_pebble-isolation_2018,
	title = {Pebble-isolation mass: {Scaling} law and implications for the formation of super-{Earths} and gas giants},
	volume = {612},
	issn = {0004-6361, 1432-0746},
	shorttitle = {Pebble-isolation mass},
	url = {https://www.aanda.org/10.1051/0004-6361/201731931},
	doi = {10.1051/0004-6361/201731931},
	language = {en},
	urldate = {2024-01-12},
	journal = {Astronomy \& Astrophysics},
	author = {Bitsch, Bertram and Morbidelli, Alessandro and Johansen, Anders and Lega, Elena and Lambrechts, Michiel and Crida, Aurélien},
	month = apr,
	year = {2018},
	pages = {A30},
}

@article{lambrechts_separating_2014,
	title = {Separating gas-giant and ice-giant planets by halting pebble accretion},
	volume = {572},
	issn = {0004-6361, 1432-0746},
	url = {http://www.aanda.org/10.1051/0004-6361/201423814},
	doi = {10.1051/0004-6361/201423814},
	language = {en},
	urldate = {2024-01-12},
	journal = {Astronomy \& Astrophysics},
	author = {Lambrechts, M. and Johansen, A. and Morbidelli, A.},
	month = dec,
	year = {2014},
	pages = {A35},
}

@article{fung_gap_2016,
	title = {{GAP} {OPENING} {IN} {3D}: {SINGLE}-{PLANET} {GAPS}},
	volume = {832},
	issn = {1538-4357},
	shorttitle = {{GAP} {OPENING} {IN} {3D}},
	url = {https://iopscience.iop.org/article/10.3847/0004-637X/832/2/105},
	doi = {10.3847/0004-637X/832/2/105},
	language = {en},
	number = {2},
	urldate = {2023-11-01},
	journal = {The Astrophysical Journal},
	author = {Fung, Jeffrey and Chiang, Eugene},
	month = nov,
	year = {2016},
	pages = {105},
}

@article{szulagyi_accretion_2014,
	title = {{ACCRETION} {OF} {JUPITER}-{MASS} {PLANETS} {IN} {THE} {LIMIT} {OF} {VANISHING} {VISCOSITY}},
	volume = {782},
	issn = {0004-637X, 1538-4357},
	url = {https://iopscience.iop.org/article/10.1088/0004-637X/782/2/65},
	doi = {10.1088/0004-637X/782/2/65},
	language = {en},
	number = {2},
	urldate = {2023-11-01},
	journal = {The Astrophysical Journal},
	author = {Szulágyi, J. and Morbidelli, A. and Crida, A. and Masset, F.},
	month = jan,
	year = {2014},
	pages = {65},
}

@article{szulagyi_meridional_2022,
	title = {Meridional {Circulation} of {Dust} and {Gas} in the {Circumstellar} {Disk}: {Delivery} of {Solids} onto the {Circumplanetary} {Region}},
	volume = {924},
	issn = {0004-637X, 1538-4357},
	shorttitle = {Meridional {Circulation} of {Dust} and {Gas} in the {Circumstellar} {Disk}},
	url = {https://iopscience.iop.org/article/10.3847/1538-4357/ac32d1},
	doi = {10.3847/1538-4357/ac32d1},
	language = {en},
	number = {1},
	urldate = {2023-10-31},
	journal = {The Astrophysical Journal},
	author = {Szulágyi, J. and Binkert, F. and Surville, C.},
	month = jan,
	year = {2022},
	pages = {1},
}

@article{hunter_matplotlib_2007,
	title = {Matplotlib: {A} {2D} graphics environment},
	volume = {9},
	doi = {10.1109/MCSE.2007.55},
	number = {3},
	journal = {Computing in Science \& Engineering},
	author = {Hunter, J. D.},
	year = {2007},
	note = {Publisher: IEEE COMPUTER SOC},
	pages = {90--95},
}

@article{harris_array_2020,
	title = {Array programming with {NumPy}},
	volume = {585},
	url = {https://doi.org/10.1038/s41586-020-2649-2},
	doi = {10.1038/s41586-020-2649-2},
	number = {7825},
	journal = {Nature},
	author = {Harris, Charles R. and Millman, K. Jarrod and Walt, Stéfan J. van der and Gommers, Ralf and Virtanen, Pauli and Cournapeau, David and Wieser, Eric and Taylor, Julian and Berg, Sebastian and Smith, Nathaniel J. and Kern, Robert and Picus, Matti and Hoyer, Stephan and Kerkwijk, Marten H. van and Brett, Matthew and Haldane, Allan and Río, Jaime Fernández del and Wiebe, Mark and Peterson, Pearu and Gérard-Marchant, Pierre and Sheppard, Kevin and Reddy, Tyler and Weckesser, Warren and Abbasi, Hameer and Gohlke, Christoph and Oliphant, Travis E.},
	month = sep,
	year = {2020},
	note = {Publisher: Springer Science and Business Media LLC},
	pages = {357--362},
}

@article{stammler_leaky_2023,
	title = {Leaky dust traps: {How} fragmentation impacts dust filtering by planets},
	volume = {670},
	issn = {0004-6361, 1432-0746},
	shorttitle = {Leaky dust traps},
	url = {https://www.aanda.org/10.1051/0004-6361/202245512},
	doi = {10.1051/0004-6361/202245512},
	language = {en},
	urldate = {2023-09-29},
	journal = {Astronomy \& Astrophysics},
	author = {Stammler, Sebastian Markus and Lichtenberg, Tim and Drążkowska, Joanna and Birnstiel, Tilman},
	month = feb,
	year = {2023},
	pages = {L5},
}

@article{masset_fargo_2000,
	title = {{FARGO}: {A} fast eulerian transport algorithm for differentiallyrotating disks},
	volume = {141},
	issn = {0365-0138, 1286-4846},
	shorttitle = {{FARGO}},
	url = {http://aas.aanda.org/10.1051/aas:2000116},
	doi = {10.1051/aas:2000116},
	language = {en},
	number = {1},
	urldate = {2023-01-12},
	journal = {Astronomy and Astrophysics Supplement Series},
	author = {Masset, F.},
	month = jan,
	year = {2000},
	pages = {165--173},
}

@article{krijt_tracing_2016,
	title = {{TRACING} {WATER} {VAPOR} {AND} {ICE} {DURING} {DUST} {GROWTH}},
	volume = {833},
	issn = {1538-4357},
	url = {https://iopscience.iop.org/article/10.3847/1538-4357/833/2/285},
	doi = {10.3847/1538-4357/833/2/285},
	language = {en},
	number = {2},
	urldate = {2022-07-30},
	journal = {The Astrophysical Journal},
	author = {Krijt, Sebastiaan and Ciesla, Fred J. and Bergin, Edwin A.},
	month = dec,
	year = {2016},
	pages = {285},
}

@article{van_clepper_chemical_2022,
	title = {Chemical {Feedback} of {Pebble} {Growth}: {Impacts} on {CO} depletion and {C}/{O} ratios},
	volume = {927},
	issn = {0004-637X, 1538-4357},
	shorttitle = {Chemical {Feedback} of {Pebble} {Growth}},
	url = {https://iopscience.iop.org/article/10.3847/1538-4357/ac511b},
	doi = {10.3847/1538-4357/ac511b},
	language = {en},
	number = {2},
	urldate = {2022-05-24},
	journal = {The Astrophysical Journal},
	author = {Van Clepper, Eric and Bergner, Jennifer B. and Bosman, Arthur D. and Bergin, Edwin and Ciesla, Fred J.},
	month = mar,
	year = {2022},
	pages = {206},
}

@article{youdin_particle_2007,
	title = {Particle stirring in turbulent gas disks: {Including} orbital oscillations},
	volume = {192},
	issn = {00191035},
	shorttitle = {Particle stirring in turbulent gas disks},
	url = {https://linkinghub.elsevier.com/retrieve/pii/S0019103507003181},
	doi = {10.1016/j.icarus.2007.07.012},
	language = {en},
	number = {2},
	urldate = {2022-05-05},
	journal = {Icarus},
	author = {Youdin, Andrew N. and Lithwick, Yoram},
	month = dec,
	year = {2007},
	pages = {588--604},
}

@article{rafikov_planet_2002,
	title = {Planet {Migration} and {Gap} {Formation} by {Tidally} {Induced} {Shocks}},
	volume = {572},
	issn = {0004-637X, 1538-4357},
	url = {https://iopscience.iop.org/article/10.1086/340228},
	doi = {10.1086/340228},
	language = {en},
	number = {1},
	urldate = {2022-05-03},
	journal = {The Astrophysical Journal},
	author = {Rafikov, R. R.},
	month = jun,
	year = {2002},
	pages = {566--579},
}

@article{zhang_systematic_2019,
	title = {Systematic {Variations} of {CO} {Gas} {Abundance} with {Radius} in {Gas}-rich {Protoplanetary} {Disks}},
	volume = {883},
	issn = {1538-4357},
	url = {https://iopscience.iop.org/article/10.3847/1538-4357/ab38b9},
	doi = {10.3847/1538-4357/ab38b9},
	language = {en},
	number = {1},
	urldate = {2022-04-29},
	journal = {The Astrophysical Journal},
	author = {Zhang, Ke and Bergin, Edwin A. and Schwarz, Kamber and Krijt, Sebastiaan and Ciesla, Fred},
	month = sep,
	year = {2019},
	pages = {98},
}

@article{birnstiel_dust_2011,
	title = {Dust size distributions in coagulation/fragmentation equilibrium: numerical solutions and analytical fits},
	volume = {525},
	issn = {0004-6361, 1432-0746},
	shorttitle = {Dust size distributions in coagulation/fragmentation equilibrium},
	url = {http://www.aanda.org/10.1051/0004-6361/201015228},
	doi = {10.1051/0004-6361/201015228},
	language = {en},
	urldate = {2022-04-29},
	journal = {Astronomy \& Astrophysics},
	author = {Birnstiel, T. and Ormel, C. W. and Dullemond, C. P.},
	month = jan,
	year = {2011},
	pages = {A11},
}

@article{drazkowska_including_2019,
	title = {Including {Dust} {Coagulation} in {Hydrodynamic} {Models} of {Protoplanetary} {Disks}: {Dust} {Evolution} in the {Vicinity} of a {Jupiter}-mass {Planet}},
	volume = {885},
	issn = {1538-4357},
	shorttitle = {Including {Dust} {Coagulation} in {Hydrodynamic} {Models} of {Protoplanetary} {Disks}},
	url = {https://iopscience.iop.org/article/10.3847/1538-4357/ab46b7},
	doi = {10.3847/1538-4357/ab46b7},
	language = {en},
	number = {1},
	urldate = {2022-04-28},
	journal = {The Astrophysical Journal},
	author = {Drążkowska, Joanna and Li, Shengtai and Birnstiel, Til and Stammler, Sebastian M. and Li, Hui},
	month = nov,
	year = {2019},
	pages = {91},
}

@article{schneider_how_2021,
	title = {How drifting and evaporating pebbles shape giant planets: {II}. {Volatiles} and refractories in atmospheres},
	volume = {654},
	issn = {0004-6361, 1432-0746},
	shorttitle = {How drifting and evaporating pebbles shape giant planets},
	url = {https://www.aanda.org/10.1051/0004-6361/202141096},
	doi = {10.1051/0004-6361/202141096},
	language = {en},
	urldate = {2022-04-29},
	journal = {Astronomy \& Astrophysics},
	author = {Schneider, Aaron David and Bitsch, Bertram},
	month = oct,
	year = {2021},
	pages = {A72},
}

@article{schneider_how_2021-1,
	title = {How drifting and evaporating pebbles shape giant planets: {I}. {Heavy} element content and atmospheric {C}/{O}},
	volume = {654},
	issn = {0004-6361, 1432-0746},
	shorttitle = {How drifting and evaporating pebbles shape giant planets},
	url = {https://www.aanda.org/10.1051/0004-6361/202039640},
	doi = {10.1051/0004-6361/202039640},
	language = {en},
	urldate = {2022-04-29},
	journal = {Astronomy \& Astrophysics},
	author = {Schneider, Aaron David and Bitsch, Bertram},
	month = oct,
	year = {2021},
	pages = {A71},
}

@article{barnett_thermal_2022,
	title = {Thermal {Processing} of {Solids} {Encountering} a {Young} {Jovian} {Core}},
	volume = {925},
	issn = {0004-637X, 1538-4357},
	url = {https://iopscience.iop.org/article/10.3847/1538-4357/ac4417},
	doi = {10.3847/1538-4357/ac4417},
	language = {en},
	number = {2},
	urldate = {2022-04-29},
	journal = {The Astrophysical Journal},
	author = {Barnett, Megan N. and Ciesla, Fred J.},
	month = feb,
	year = {2022},
	pages = {141},
}

@article{morbidelli_meridional_2014,
	title = {Meridional circulation of gas into gaps opened by giant planets in three-dimensional low-viscosity disks},
	volume = {232},
	url = {krakr},
	doi = {10.1016/j.icarus.2014.01.010},
	language = {en},
	number = {1},
	urldate = {2022-04-28},
	journal = {Icarus},
	author = {Morbidelli, A. and Szulágyi, J and Crida, A. and Lega, E. and Bitsch, B. and Tanigawa, T. and Kanagawa, K.},
	year = {2014},
	note = {arXiv: 1309.2871},
	pages = {266--270},
}

@article{weber_characterizing_2018,
	title = {Characterizing the {Variable} {Dust} {Permeability} of {Planet}-induced {Gaps}},
	volume = {854},
	issn = {1538-4357},
	url = {https://iopscience.iop.org/article/10.3847/1538-4357/aaab63},
	doi = {10.3847/1538-4357/aaab63},
	language = {en},
	number = {2},
	urldate = {2022-04-28},
	journal = {The Astrophysical Journal},
	author = {Weber, Philipp and Benítez-Llambay, Pablo and Gressel, Oliver and Krapp, Leonardo and Pessah, Martin E.},
	month = feb,
	year = {2018},
	pages = {153},
}

@article{haugbolle_probing_2019,
	title = {Probing the {Protosolar} {Disk} {Using} {Dust} {Filtering} at {Gaps} in the {Early} {Solar} {System}},
	volume = {158},
	issn = {1538-3881},
	url = {https://iopscience.iop.org/article/10.3847/1538-3881/ab1591},
	doi = {10.3847/1538-3881/ab1591},
	language = {en},
	number = {2},
	urldate = {2022-04-28},
	journal = {The Astronomical Journal},
	author = {Haugbølle, Troels and Weber, Philipp and Wielandt, Daniel P. and Benítez-Llambay, Pablo and Bizzarro, Martin and Gressel, Oliver and Pessah, Martin E.},
	month = jul,
	year = {2019},
	pages = {55},
}

@article{chiang_spectral_1997,
	title = {Spectral {Energy} {Distributions} of {T} {Tauri} {Stars} with {Passive} {Circumstellar} {Disks}},
	volume = {490},
	issn = {0004-637X, 1538-4357},
	url = {https://iopscience.iop.org/article/10.1086/304869},
	doi = {10.1086/304869},
	language = {en},
	number = {1},
	urldate = {2022-04-28},
	journal = {The Astrophysical Journal},
	author = {Chiang, E. I. and Goldreich, P.},
	month = nov,
	year = {1997},
	pages = {368--376},
}

@article{misener_tracking_2019,
	title = {Tracking {Dust} {Grains} during {Transport} and {Growth} in {Protoplanetary} {Disks}},
	volume = {885},
	issn = {1538-4357},
	url = {https://iopscience.iop.org/article/10.3847/1538-4357/ab4a13},
	doi = {10.3847/1538-4357/ab4a13},
	language = {en},
	number = {2},
	urldate = {2022-04-27},
	journal = {The Astrophysical Journal},
	author = {Misener, William and Krijt, Sebastiaan and Ciesla, Fred J.},
	month = nov,
	year = {2019},
	pages = {118},
}

@article{krapp_3d_2022,
	title = {The {3D} {Dust} and {Opacity} {Distribution} of {Protoplanets} in {Multifluid} {Global} {Simulations}},
	volume = {928},
	issn = {0004-637X, 1538-4357},
	url = {https://iopscience.iop.org/article/10.3847/1538-4357/ac5899},
	doi = {10.3847/1538-4357/ac5899},
	language = {en},
	number = {2},
	urldate = {2022-04-27},
	journal = {The Astrophysical Journal},
	author = {Krapp, Leonardo and Kratter, Kaitlin M. and Youdin, Andrew N.},
	month = apr,
	year = {2022},
	pages = {156},
}

@article{teague_meridional_2019,
	title = {Meridional flows in the disk around a young star},
	volume = {574},
	issn = {0028-0836, 1476-4687},
	url = {http://www.nature.com/articles/s41586-019-1642-0},
	doi = {10.1038/s41586-019-1642-0},
	language = {en},
	number = {7778},
	urldate = {2022-03-29},
	journal = {Nature},
	author = {Teague, Richard and Bae, Jaehan and Bergin, Edwin A.},
	month = oct,
	year = {2019},
	pages = {378--381},
}

@article{alarcon_chemical_2020,
	title = {Chemical {Evolution} in a {Protoplanetary} {Disk} within {Planet} {Carved} {Gaps} and {Dust} {Rings}},
	volume = {905},
	issn = {0004-637X, 1538-4357},
	url = {https://iopscience.iop.org/article/10.3847/1538-4357/abc1d6},
	doi = {10.3847/1538-4357/abc1d6},
	language = {en},
	number = {1},
	urldate = {2022-03-04},
	journal = {The Astrophysical Journal},
	author = {Alarcón, Felipe and Teague, R. and Zhang, Ke and Bergin, E. A. and Barraza-Alfaro, M.},
	month = dec,
	year = {2020},
	pages = {68},
}

@article{bergner_ice_2021,
	title = {Ice {Inheritance} in {Dynamical} {Disk} {Models}},
	volume = {919},
	issn = {0004-637X, 1538-4357},
	url = {https://iopscience.iop.org/article/10.3847/1538-4357/ac0fd7},
	doi = {10.3847/1538-4357/ac0fd7},
	language = {en},
	number = {1},
	urldate = {2022-02-11},
	journal = {The Astrophysical Journal},
	author = {Bergner, Jennifer B. and Ciesla, Fred},
	month = sep,
	year = {2021},
	pages = {45},
}

@article{oberg_effects_2011,
	title = {{THE} {EFFECTS} {OF} {SNOWLINES} {ON} {C}/{O} {IN} {PLANETARY} {ATMOSPHERES}},
	volume = {743},
	issn = {2041-8205, 2041-8213},
	url = {https://iopscience.iop.org/article/10.1088/2041-8205/743/1/L16},
	doi = {10.1088/2041-8205/743/1/L16},
	language = {en},
	number = {1},
	urldate = {2021-05-12},
	journal = {The Astrophysical Journal},
	author = {Öberg, Karin I. and Murray-Clay, Ruth and Bergin, Edwin A.},
	month = dec,
	year = {2011},
	pages = {L16},
}

@article{mcelroy_umist_2013,
	title = {The {UMIST} database for astrochemistry 2012},
	volume = {550},
	issn = {0004-6361, 1432-0746},
	url = {http://www.aanda.org/10.1051/0004-6361/201220465},
	doi = {10.1051/0004-6361/201220465},
	language = {en},
	urldate = {2021-05-06},
	journal = {Astronomy \& Astrophysics},
	author = {McElroy, D. and Walsh, C. and Markwick, A. J. and Cordiner, M. A. and Smith, K. and Millar, T. J.},
	month = feb,
	year = {2013},
	pages = {A36},
}

@article{benitez-llambay_fargo3d_2016,
	title = {{FARGO3D}: {A} {NEW} {GPU}-{ORIENTED} {MHD} {CODE}},
	volume = {223},
	issn = {1538-4365},
	shorttitle = {{FARGO3D}},
	url = {https://iopscience.iop.org/article/10.3847/0067-0049/223/1/11},
	doi = {10.3847/0067-0049/223/1/11},
	language = {en},
	number = {1},
	urldate = {2021-05-05},
	journal = {The Astrophysical Journal Supplement Series},
	author = {Benítez-Llambay, Pablo and Masset, Frédéric S.},
	month = mar,
	year = {2016},
	pages = {11},
}

@article{weidenschilling_aerodynamics_1977,
	title = {Aerodynamics of solid bodies in the solar nebula},
	volume = {180},
	issn = {0035-8711, 1365-2966},
	url = {https://academic.oup.com/mnras/article-lookup/doi/10.1093/mnras/180.2.57},
	doi = {10.1093/mnras/180.2.57},
	language = {en},
	number = {2},
	urldate = {2021-05-05},
	journal = {Monthly Notices of the Royal Astronomical Society},
	author = {Weidenschilling, S. J.},
	month = sep,
	year = {1977},
	pages = {57--70},
}

@article{birnstiel_simple_2012,
	title = {A simple model for the evolution of the dust population in protoplanetary disks},
	volume = {539},
	issn = {0004-6361, 1432-0746},
	url = {http://www.aanda.org/10.1051/0004-6361/201118136},
	doi = {10.1051/0004-6361/201118136},
	language = {en},
	urldate = {2020-11-17},
	journal = {Astronomy \& Astrophysics},
	author = {Birnstiel, T. and Klahr, H. and Ercolano, B.},
	month = mar,
	year = {2012},
	pages = {A148},
}
\bibliographystyle{aasjournalv7}

% \appendix
% \section{Appendix}
% Appendix can go here if necessary.

\end{document}